\documentclass[symmetry, article, accept, oneauthor, pdftex]{mdpi}

\usepackage{latexsym}
\usepackage{amsmath}
\usepackage{amssymb}
\usepackage{amsfonts}

\usepackage{placeins}
\usepackage{supertabular}

\usepackage[mathscr,scaled=1.15]{urwchancal}
\DeclareFontFamily{OT1}{pzc}{}
\DeclareFontShape{OT1}{pzc}{m}{it}%
{<-> s * [1.15] pzcmi7t}{}
\DeclareMathAlphabet{\mathpzc}{OT1}{pzc}{m}{it}

\firstpage{1}
\pubvolume{xx}
\issuenum{1}
\articlenumber{5}
\pubyear{2020}
\copyrightyear{2020}
\history{Received: 28 July 2020; Accepted: 28 August 2020; Published: date}

\Title{Empirical Consequences of Emergent Mass}

\newcommand{\orcidauthorA}{0000-0002-2937-1361} 

\Author{{Craig D. Roberts} $^{1,2}$}  

\AuthorNames{Craig D. Roberts\orcidauthorA{}}

\address{%
$^{1}$ \quad School of Physics, Nanjing University, Nanjing 210093, Jiangsu, China; cdroberts@nju.edu.cn\\
$^{2}$ \quad Institute for Nonperturbative Physics, Nanjing University, Nanjing  210093, Jiangsu, China} %

\corres{Correspondence: cdroberts@nju.edu.cn}

\abstract{The Lagrangian that defines quantum chromodynamics (QCD), the~strong interaction piece of the Standard Model, appears very simple.  Nevertheless, it is responsible for an astonishing array of high-level phenomena with enormous apparent complexity, e.g., the~existence, number and structure of atomic nuclei.  The source of all these things can be traced to emergent mass, which~might itself be QCD's self-stabilising mechanism.  A background to this perspective is provided, presenting, inter alia, a discussion of the gluon mass and QCD's process-independent effective charge and highlighting an array of observable expressions of emergent mass, ranging from its manifestations in pion parton distributions to those in nucleon electromagnetic form factors.}

\begin{document}



\section{Introduction}
One might define \emph{emergent phenomena} as those features of nature which do not readily admit an explanation solely in terms of known or conjectured mathematical rules.  The~concept is at least as old as Aristotle (384--322\,BC), who argued that a compound item can have (emergent) properties in the whole which are not explicable merely through the independent actions of the item's constituent parts.  His view is often represented by the statement ``The whole is more than the sum of its parts.''  In this sense, \emph{emergence} has its origins in the Greek ``sunergos'': ``together'' plus ``working'', which~is the origin of our current concept of synergy, viz.\ things working \emph{together} more effectively than could be anticipated from their independent actions in isolation. (Etymologically, the~word ``emergence'' entered English in the mid 17th century, meaning ``unforeseen occurrence'' and derived from the medieval Latin ``emergentia'', itself from the Latin ``emergere'', meaning ``bring to light''  (Source: Oxford English Dictionary).  Herein, ``emergence'' is seen as a larger notion than is typically expected from this literal connection.).

This perspective is typically contrasted with that described as \emph{reductionism}; namely, the~view that everything in nature can ultimately be viewed as no more complex in principle than, e.g.,\ a (very~good) watch, which~is clearly a complex object; but, equally clearly, not more than the sum of its~parts.

In developing such a contrast through debate, hard lines are sometimes drawn, with~individuals deciding or being forced to choose one side or the other.  This~is taking the argument too far, however.  During~each epoch in history, there has always been a line dividing physics from metaphysics; but~the location of that border is neither fixed nor impermeable.  As~time and humanity have progressed, more~aspects of nature have seeped into the pool of physics.  Notwithstanding such progress, it does not follow that fewer questions have been left beyond physics.  Typically, as~mathematics has succeeded in explaining more phenomena, new discoveries have been made, often emerging from attempts to test the newly formulated theories.  (One may think here of the discovery of the neutron and then the plethora of other so-called elementary particles, to~which the introduction of order demanded the development of quantum field theory.)  Therefore, the~tension remains.  If~nature is bounded, then this might change at some future time; but today it would be gross hubris to maintain such a~position.

The question ``Can all objects that have emerged in nature be explained by a finite collection of rules?'' is at least as old as human thought, and we cannot know the answer until all that nature can produce has been discovered.  Plainly, the~debate must continue, and, perhaps, its greatest merit is the spur that each side provides the other as we seek to understand our place in \emph{the scheme of things}.

\section{Strong Interactions in the Standard~Model}
\unskip
\subsection{Natural Mass~Scale}
A significant part of the ongoing debate centres on the character of mass and its consequences in the Standard Model of Particle Physics (SM), especially as it emerges from the strong interaction sector; namely, quantum chromodynamics (QCD).  To~introduce this problem, it is worth recalling the Nobel Prize acceptance speech given by H.\,D.~Politzer; in particular, the~following remarks~\cite{Politzer:2005kc}.
\begin{quote}
\emph{The establishment by the mid-1970's of QCD as the correct theory of the strong interactions completed what is now known prosaically as the Standard Model.  It offers a description of all known fundamental physics except for gravity, and~gravity is something that has no discernible effect when particles are studied a few at a time.  However, the~situation is a bit like the way that the Navier-Stokes equation accounts for the flow of water.  The~equations are at some level obviously correct, but~there are only a few, limited circumstances in which their consequences can be worked out in any detail.  Nevertheless, many leading physicists were inclined to conclude in the late 1970's that the task of basic physics was nearly complete, and~we'd soon be out of jobs.  A~famous example was the inaugural lecture of Stephen Hawking as Lucasian Professor of Mathematics, a~chair first held by Isaac Barrow at Cambridge University.  Hawking titled his lecture, `Is the End in Sight for Theoretical Physics?'  And he argued strongly for `Yes'}.
\end{quote}

Concerning the character of mass, many might believe that the answer was found in 2012 with the discovery of the Higgs boson~\cite{Aad:2012tfa, Chatrchyan:2012xdj} and the subsequent Nobel Prize awarded in equal share to F.~Englert and P.~Higgs~\cite{Englert:2014zpa, Higgs:2014aqa}, with~the citation~``for the theoretical discovery of a mechanism that contributes to our understanding of the origin of mass of subatomic particles \ldots''.~Nevertheless,~while discovery of the Higgs was a watershed, it should be placed in context; something achieved nicely and informally on The Guardian's live blog: %
(\href{https://www.theguardian.com/science/2011/dec/13/higgs-boson-seminar-god-particle}
{theguardian.com/science/2011/dec/13/higgs-boson-seminar-god-particle}
\begin{quote}
\emph{The Higgs field is often said to give mass to everything.  That is wrong.  The~Higgs field only gives mass to some very simple particles.  The~field accounts for only one or two percent of the mass of more complex things, like atoms, molecules, and everyday objects, from~your mobile phone to your pet llama. The~vast majority of mass comes from the energy needed to hold quarks together inside atoms}.
\end{quote}

These remarks implicitly highlight QCD, the~quantum field theory formulated in four spacetime dimensions which defines what is arguably the SM's most important chapter.  QCD is supposed to describe all of nuclear physics through the interactions between quarks (matter fields) that are mediated by gluons (gauge bosons).  Yet, fifty years after the discovery of quarks~\cite{Taylor:1991ew, Kendall:1991np, Friedman:1991nq}, science is only just beginning to grasp how QCD moulds the basic bricks for nuclei: pions, neutrons, protons, etc., and it is far from understanding how QCD produces~nuclei.

The natural scale for nuclear physics (strong interactions) is characterised by the proton mass:
\begin{equation}
m_p \approx 1\,{\rm GeV} \approx 2000\,m_e\,,
\end{equation}
where $m_e$ is the electron mass, i.e.,\ $m_p = 1.783 \times 10^{-27}\,$kg.  In~the SM, $m_e$ is correctly attributed to the Higgs boson, but what is the cause of the prodigious enhancement required to produce~$m_p$?  Followed~logically to its source, this question leads to an appreciation that our Universe's existence depends critically on, inter alia, the~following empirical facts.
(\emph{i}) The proton mass is large, i.e.,\ the~mass-scale for QCD is very much larger than that of electromagnetism;
(\emph{ii}) the proton does not decay, despite being a compound state built from three valence-quarks;
and (\emph{iii}) the pion, which~carries long-range interactions between nucleons (neutrons and protons), is abnormally light (not massless), having a lepton-like mass even though it is a strongly interacting object built from a valence-quark and valence antiquark. (The $\mu$-lepton, discovered in 1936~\cite{PhysRev.51.884}, was initially mistaken for the pion.  The~pion was only found a decade later~\cite{Lattes:1947mw}.)
These qualities of Nature transport us to a reductionist definition of emergence.  Namely, assuming it does describe strong interactions, then the one-line Lagrangian of QCD---a very simple low-level rule---must somehow produce high-level phenomena with enormous apparent~complexity.

At this point, it is worth studying the Lagrangian of chromodynamics, which~appeared as the culmination of a distillation process applied to a large array of distinct ideas and discoveries~\cite{Marciano:1977su, Marciano:1979wa}:
\begin{subequations}
\label{QCDdefine}
\begin{align}
{\mathpzc L}_{\rm QCD} & = \sum_{j=u,d,s,\ldots}
\bar{q}_j [\gamma_\mu D_{\mu} + m_j] q_j + \tfrac{1}{4} G^a_{\mu\nu} G^{a\mu\nu},\\
D_{\mu} & = \partial_\mu + i g \tfrac{1}{2} \lambda^a A^a_\mu\,, \quad
\label{gluonSI}
G^a_{\mu\nu} = \partial_\mu A^a_\nu + \partial_\nu A^a_\mu -
\underline{{g f^{abc}A^b_\mu A^c_\nu}}.
\end{align}
\end{subequations}

Here, $\{q_j\}$ are the quark fields, with~$j$ their flavour label and $m_j$ their Higgs-generated current-quark masses, and $\{ A_\mu^a, a=1,\ldots, 8 \}$ are the gluon fields, with~$\{\tfrac{1}{2} \lambda^a\}$ the generators of the SU$(3)$ (chromo/colour) gauge-group in the fundamental representation.~Comparing with quantum electrodynamics (QED), the~solitary difference is the piece describing gluon self-interactions, marked~as the underlined term in Equation\,\eqref{gluonSI}.  Somehow, the~origin, mass and~extent of almost all visible matter in the Universe is attributable to ${\mathpzc L}_{\rm QCD}$---one line plus two definitions.  That being true, then \ldots
\begin{quote}
\emph{QCD is quite possibly the most remarkable fundamental theory ever invented.}
\end{quote}

The only apparent energy scales in Equation\,\eqref{QCDdefine} are the Higgs-generated current-quark masses, but focusing on the $u$ (up) and $d$ (down) quarks that define nucleons, this scale is more-than one-hundred-times smaller than $m_p$ \cite{Zyla:2020}.  No amount of ``staring'' at ${\mathpzc L}_{\rm QCD}$ can reveal the source of that enormous amount of ``missing mass''; yet, it must be there. (This is a stark contrast to QED wherein, e.g.,\ the~scale in the spectrum of the hydrogen atom is set by $m_e$, which~is a prominent feature of ${\mathpzc L}_{\rm QED}$ that is generated by the Higgs boson.).

Models and effective field theories (EFTs) for nuclear physics typically assume existence of the $m_p \approx 1\,$GeV mass-scale and build upon it.  They also assume the reality of effectively pointlike nuclear constituents (nucleons) and force carriers (pions and, perhaps, other meson-like entities).  Their task is not to elucidate the internal structure of such objects.  Instead, they aim to develop systematically improvable techniques that can describe the number and properties of atomic nuclei.  This~may be seen as reductionism built on an emergent plateau.  The~basic reductionist question generated by this approach is ``Can the plateau upon which the nuclear model/EFT paradigm is built be constructed from QCD?''  If the answer is ``yes'', then all parameters used and fitted in such theories of nuclear structure will (some day) be confronted with ab~initio predictions in a profound test of the~SM.

\begin{figure}[t]
\begin{center}
\includegraphics[clip,width=0.60\linewidth]{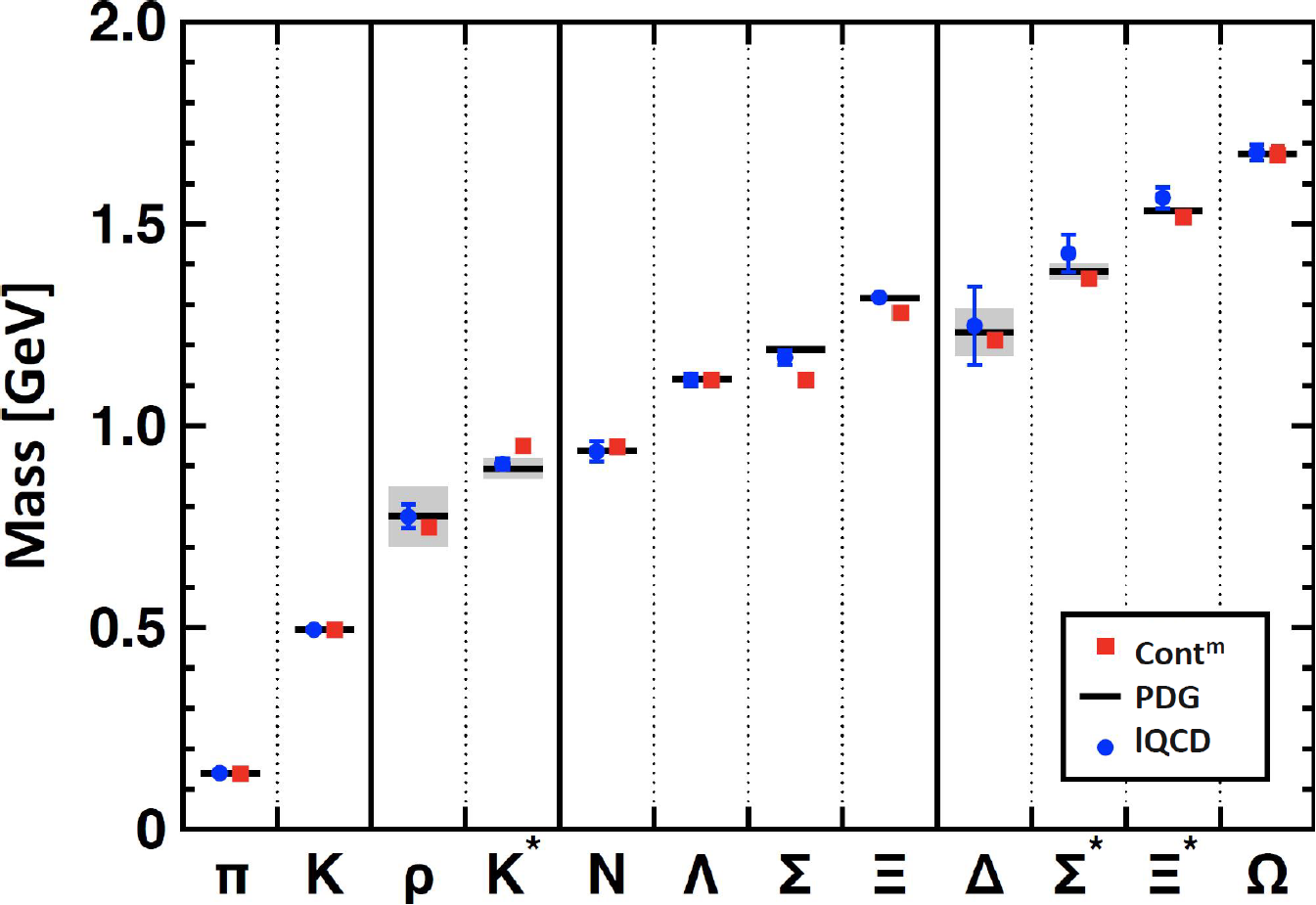}
\end{center}
\caption{
Masses of pseudoscalar and vector mesons, and~ground-state positive-parity octet and decuplet baryons calculated using continuum (Cont$^{\rm m}$---squares, red) \cite{Qin:2019hgk} and lattice (lQCD---circles, blue) \cite{Durr:2008zz} methods in QCD compared with experiment~\cite{Zyla:2020} (PDG---black bars, with~decay-widths of unstable states shaded in grey).  The~continuum study did not include isospin symmetry breaking effects, which~are evidently small, as~highlighted by the empirically determined $\Sigma$-$\Lambda$ mass difference~($<$7\%).
}
\label{810MassComparison}
\end{figure}

Today there is a good case to be made for an affirmative answer to this question.  Amongst the numerous supporting examples, one may list QCD-connected computations of the hadron spectrum. Depicted in Figure\,\ref{810MassComparison}, one sees that two disparate approaches to solving QCD~\cite{Qin:2019hgk, Durr:2008zz} produce a spectrum of ground-state hadrons in good agreement with~experiment.

Importantly, however, neither calculation presented in Figure\,\ref{810MassComparison} was able to predict the size of the proton mass.  Apart from the Higgs-generated current-quark masses, each has an undetermined mass-scale parameter, denoted hereafter as $\Lambda_{\rm QCD}$.  Its value is chosen to fit one experimental number, e.g.,\ the~pion's leptonic decay constant, but after $\Lambda_{\rm QCD}$ is fixed, all other results are~predictions.

\subsection{Whence Mass?}
\label{secWhence}
Herein, it is worth explaining the need for such a scale-setting procedure.
Regarded classically, chromodynamics is a local non-Abelian gauge field theory, and defined in four spacetime dimensions, there is no mass-scale if Lagrangian masses for the fermions are omitted.  (The absence of such masses defines the chiral limit.)  Scale invariant theories do not support dynamics, only kinematics.  Therefore, bound states are impossible; so, the~Universe cannot~exist.

Spontaneous symmetry breaking, as~generated by the Higgs mechanism, does not resolve this issue.  The~masses of the neutron and proton, which~lie at the heart of all visible matter, are two orders-of-magnitude greater than the Higgs-generated current-masses of the $u$- and $d$-quarks, which~are the defining constituents of protons and~neutrons.

Consequently, questions like ``How does a mass-scale appear?'' and ``Why does this scale have its observed value?''  are consanguineous with the question ``How did the Universe form?"

Modern quantum field theories are built upon Poincar\'e invariance.  In~this connection, consider the energy-momentum tensor in classical chromodynamics, $T_{\mu\nu}$.  In~field theory, conservation of energy and momentum is a consequence of spacetime translational invariance, one of the family of Poincar\'e transformations.  Consequently,
\begin{equation}
\label{ConsEP}
\partial_\mu T_{\mu\nu} = 0 \,.
\end{equation}

Consider now a global scale transformation in the Lagrangian of classical chromodynamics:
\begin{equation}
\label{scaleT}
x \to x^\prime = {\rm e}^{-\sigma}x\,, \quad
A_\mu^a(x) \to A_\mu^{a\prime}(x^\prime) = {\rm e}^{-\sigma} A_\mu^a({\rm e}^{-\sigma}x ) \,,\quad
q(x)  \to q^\prime(x^\prime) = {\rm e}^{- (3/2) \sigma} q({\rm e}^{-\sigma}x )\,,
\end{equation}
where $\sigma$ is spacetime-independent.  The~associated Noether current is
\begin{equation}
{\cal D}_\mu = T_{\mu\nu} x_\nu\,,
\end{equation}
viz.\ the dilation current.  In~the absence of fermion masses, the~classical action is invariant under Equation\,\eqref{scaleT}, i.e.,\ the~theory is scale invariant; therefore,
\begin{equation}
\partial_\mu {\cal D}_\mu  = 0  = [\partial_\mu T_{\mu\nu} ] x_\nu + T_{\mu\nu} \delta_{\mu\nu}  = T_{\mu\mu}\,,
\label{SIcQCD}
\end{equation}
where the last equality follows from Equation\,\eqref{ConsEP}.  Evidently, the~energy-momentum tensor must be traceless in a scale invariant~theory.

Classical chromodynamics is not a meaningful physical framework for many reasons; amongst~them the fact, illustrated by Figure\,\ref{810MassComparison}, that strong interactions are empirically known to be characterised by a large mass-scale, $m_p \approx 1\,$GeV.  In~quantising the theory, a~mass-scale is introduced by regularisation and renormalisation of ultraviolet divergences.  This~is ``dimensional transmutation'': all quantities, including the field operators themselves, become dependent on a mass-scale.  This~entails the violation of Equation\,\eqref{SIcQCD}, i.e.,\ the~appearance of the chiral-limit ``trace anomaly'' \cite{tarrach}:
\begin{equation}
\label{SIQCD}
T_{\mu\mu} = \beta(\alpha(\zeta))  \tfrac{1}{4} G^{a}_{\mu\nu}G^{a}_{\mu\nu} =: \Theta_0 \,,
\end{equation}
where $\beta(\alpha(\zeta))$ is QCD's $\beta$-function, $\alpha(\zeta)$ is the associated running-coupling, and~$\zeta$ is the renormalisation scale.  Equation \eqref{SIQCD} indicates that a mass-scale linked to the resolving power of a given measurement is engendered via quantisation; to wit, the~scale \emph{emerges} as an indispensable part of the theory's quantum~definition.

This mass is exhibited in the gauge-boson vacuum polarization. In~QED, the~photon vacuum polarization does not possess an infrared mass-scale, and~dimensional transmutation acts simply to generate the gentle evolution of the QED coupling, i.e.,\ any dynamical breaking of QED's conformal features is small; therefore, the~associated trace anomaly is normally negligible.  In~contrast, gauge sector dynamics drives a Schwinger mechanism in QCD~\cite{Cornwall:1981zr, Gribov:1999ui, Dudal:2003by, Bowman:2004jm, Luna:2005nz, Aguilar:2008xm, RodriguezQuintero:2010wy, Boucaud:2011ugS, Strauss:2012dg, Binosi:2014aea, Aguilar:2015bud, Siringo:2015wtx, Cyrol:2016tym, Gao:2017uox, Binosi:2019ecz, Binosi:2016nme, Rodriguez-Quintero:2018wma, Cui:2019dwvS}, so that the QCD trace anomaly expresses a mass-scale which is, empirically, very significant.  This~is discussed in Section\,\ref{secGluonMass}.

There is another aspect of chromodynamics that should be highlighted, namely, the~classical Lagrangian still defines a non-Abelian local gauge theory.  Accordingly, the~concept of local gauge invariance persists; but without a mass-scale, there is no notion of confinement.  For~instance, one can compose a colour-singlet combination of three quarks and colour rotations will preserve its colour neutrality; but the participating quarks need not be close together.  In~fact, it is meaningless to discuss proximity because, in~a scale invariant theory, all lengths are equivalent.  Accordingly, the~question of ``Whence mass?'' is indistinguishable from ``Whence a mass-scale?'', is indistinguishable from ``Whence~a confinement scale?''.

Evidently, one does not learn much from knowing that a trace anomaly exists.  It only means there is a mass-scale.  The~central concern is whether the magnitude of that scale can be computed and~understood.

The magnitude of the scale anomaly can definitely be measured, and this simply by considering the in-proton expectation value of the energy-momentum tensor (see, e.g., in \cite{Kharzeev:1995ij}):
\begin{equation}
\label{EPTproton}
\langle p(P) | T_{\mu\nu} | p(P) \rangle = - P_\mu P_\nu\,,
\end{equation}
where the equations-of-motion for a one-particle proton state were used to obtain the right-hand-side.  Now, in~the chiral limit
\begin{equation}
\label{anomalyproton}
\langle p(P) | T_{\mu\mu} | p(P) \rangle  = - P^2  = m_p^2 = \langle p(P) |  \Theta_0 | p(P) \rangle\,.
\end{equation}
Therefore, there is a sound position from which one may conclude that gluons generate all the proton's mass: the measured value of the trace anomaly is large; and, logically, that feature owes to gluon self-interactions, the~agent behind asymptotic~freedom.

This is a valid deduction because, ultimately, what else could account for a mass-scale in QCD?  Gluon self-interactions are QCD's definitive feature, and it is these interactions that might enable a rigorous (non-perturbative) definition of the matrix element in Equation\,\eqref{anomalyproton}.  Nevertheless, it is only reasonable to conclude this when the operator and wave function are computed at a resolving scale $\zeta \gg m_p$, i.e.,\ when one employs a parton-model basis~\cite{Close:1979zpo}.

There is another issue, too, which~can be exposed by returning to Equation\,\eqref{EPTproton} and replacing the proton by the pion
\begin{equation}
\label{EPTpion}
\langle \pi(q) | T_{\mu\nu} | \pi(q) \rangle = - q_\mu q_\nu \quad  \Rightarrow \quad
 \langle \pi(q) |  \Theta_0 | \pi (q) \rangle = m_\pi^2 \;\; \stackrel{\mbox{\rm chiral\;limit}}{=} \;\; 0
\end{equation}
because the chiral-limit pion is a massless Nambu--Goldstone (NG) mode~\cite{Nambu:1960tm, Goldstone:1961eq}.
It is possible that Equation\,\eqref{EPTpion} means the scale anomaly is trivially zero in the pion, i.e.,\ gluon self-interactions have no effect in the pion because each term required to express the operator vanishes separately.  Yet, such~a conclusion would sit uncomfortably with known QCD dynamics, which~expresses both attraction and repulsion, but~never passive inactivity.
More likely, then, the~final identity in Equation\,\eqref{EPTpion} results from cancellations between different terms in the complete operator matrix element.  Naturally, such~precise cancellation could not be accidental.  It would require that some symmetry is broken in a very particular manner.  (The mechanism is explained in Section\,\ref{secPionTrace}.)

Equations\,\eqref{anomalyproton} and \eqref{EPTpion} present a quandary.  They stress that any understanding of the proton's mass is incomplete unless it simultaneously explains Equation\,\eqref{EPTpion}. Moreover, any discussion of confinement, fundamental to the proton's stability, is unreasonable before this conundrum is resolved.  As~will become clear, at~least some of these features of Nature have a reductionist explanation grounded in the dynamics responsible for the emergence of $m_p$ as the natural mass-scale for nuclear physics; and one of the most important goals in modern science is to explain and elucidate the entire array of empirical consequences of this dynamics~\cite{Aguilar:2019teb, Roberts:2019wov, Brodsky:2020vco, Roberts:2020udq, Aznauryan:2012baS}.

\section{Confinement}
\label{secConfinement}
It is a textbook result, with~its origin in the Nobel Prize for the discovery of asymptotic \mbox{freedom~\cite{Politzer:2005kc, Gross:2005kv, Wilczek:2005az}}, that QCD is characterised by an interaction which becomes stronger as the participants try to separate.  Remaining within those results that can be established using perturbation theory, one is led to contemplate some unusual possibilities: if the coupling strength rises rapidly with separation, then perhaps it is not bounded, and perhaps an infinite amount of energy is required to remove a gluon or quark from the interior of a hadron?  Such thinking has~produced
\begin{quote}
\textit{The Confinement Hypothesis}: Colour-charged particles cannot be isolated and therefore cannot be directly observed.  They clump together in colour-neutral bound-states.
\end{quote}

Confinement seems to be an empirical reality, but a credited mathematical proof is lacking.  Partly~as a result, the~Clay Mathematics Institute proffered a ``Millennium Problem'' prize of \$1 million for a proof that pure-glue QCD is mathematically well defined~\cite{Jaffe:Clay}.  One necessary part of such a proof would
be to establish whether the confinement postulate is correct in pure-gauge QCD
i.e.,\ the~theory obtained from Equation\,\eqref{QCDdefine} after omitting all terms containing quark~fields.

There is a pitfall here, however: no reader of this material can be described within pure-glue QCD.  Light quarks are essential to understanding all known visible matter.  Thus, a~proof of confinement in pure-glue QCD is chiefly irrelevant to our Universe.  Life exists because nature has supplied two light quark flavours, combinations of which form the pion, and the pion is unnaturally light, thus~very easily produced and capable of propagating over nuclear-size distances.  Therefore, as~noted previously by others~\cite{Casher:1979vw, Banks:1979yr}, no explanation of SM confinement is empirically relevant unless it also describes the link  between confinement and the emergence of mass, and~so the existence and role of pions, i.e.,\ pseudo-NG modes with $m_\pi\ll m_p$.

One piece of the Yang--Mills millennium problem~\cite{Jaffe:Clay} is to prove that a mass-gap, $\Delta>0$, exists~in pure-glue QCD.  This~conjecture is supported by some strong evidence, e.g.,\ numerical studies of lattice-regularised QCD (lQCD) find $\Delta \gtrsim 1.5\,$GeV~\cite{McNeile:2008sr}.  This~sharpens the conundrum described above: with $\Delta^2/m_\pi^2 \gtrsim 100$, can the mass-gap in pure Yang--Mills theory really play any role in understanding confinement when the emergence of mass, driven by kindred dynamics, ensures that our Universe supports almost-massless strongly-interacting excitations?
Skirting the question, one can respond that any explanation of confinement must simultaneously describe its link to pion properties.
From~this position, pions are viewed as playing a critical role in any explanation of SM confinement, and a discussion that omits reference to pions is \emph{practically} \emph{irrelevant}.

These observations indicate that the potential between infinitely-heavy quarks computed in numerical simulations of quenched lQCD---the static potential~\cite{Wilson:1974sk}, often associated with formation of an incredibly strong flux tube between the colour source and sink~\cite{Isgur:1983wj}---is detached from the question of confinement in our Universe.  In~fact, as light--particle annihilation and creation effects are essentially non-perturbative in QCD, it is impossible to calculate a quantum mechanical potential between two light quarks~\cite{Bali:2005fu, Prkacin:2005dc, Chang:2009ae}.  It follows that there is no discernible flux tube in a Universe with light quarks; so, the~flux tube is not the correct paradigm for~confinement.

As highlighted already, the~emergence of mass is key here.  It ensures the existence of pseudo-NG modes, and no flux tube linking a static colour source and sink can have an observable existence in the presence of these modes.  To~verify this statement, suppose that such a tube is stretched between a source and sink.  The~potential energy stored within the tube may only increase until it reaches the amount required to produce a particle--antiparticle pair of the system's pseudo-NG modes.  Simulations~of lQCD demonstrate~\cite{Bali:2005fu, Prkacin:2005dc} that the flux tube then disappears instantaneously, leaving~two separated colour-singlet systems.  The~length-scale characterising this effect is $r_{\!\not\,\sigma} \simeq (1/3)\,$fm; so,~if~any such string forms, it would disintegrate well within a hadron's~interior.

An alternative realisation associates confinement with marked
%
changes in the analytic properties of coloured propagators and vertices, driven by QCD dynamics.  That leads
such coloured $n$-point functions to violate the axiom of reflection positivity, thereby eliminating the associated excitations from the Hilbert space associated with asymptotic states~\cite{GJ81}.  This~is certainly a sufficient condition for confinement~\cite{Munczek:1983dx, Cahill:1985mh, Stingl:1985hx, Krein:1990sf, Burden:1991gd, Hawes:1993ef, Maris:1994ux, Roberts:1994dr, Bhagwat:2002tx, Roberts:2007ji, Bashir:2009fv, Strauss:2012dg, Bashir:2013zha, Qin:2013ufa, Lowdon:2015fig, Lucha:2016vte, Binosi:2016xxu, Binosi:2019ecz}.

It should be highlighted, however, that the appearance of such modifications when analysing some simplification of a given theory does not signify that the theory itself is truly confining: uncommon~spectral properties can be introduced by approximations, yielding a truncated version of a theory which expresses confinement even though the complete theory does not, see, e.g.,\ in \,\cite{Krein:1993jb, Bracco:1993cy}.  Notwithstanding exceptions like these, in~a veracious treatment of QCD the computed violation of reflection positivity by coloured functions does express~confinement.

\section{Strong~QCD}
\unskip
\subsection{Dyson--Schwinger~Equations}
The appearance and size of the natural scale for nuclear physics ($m_p \approx 1\,$GeV) and the confinement of gluons and quarks are emergent phenomena.  They are not apparent in the QCD Lagrangian, yet~they determine the character of QCD's spectrum, the~structure of bound states and~so forth.  Given~Equation\,\eqref{QCDdefine}, the~natural question to ask is whether one can understand these features reductively, i.e.,\ directly in terms of the degrees-of-freedom used to formulate QCD, or~does the complexity of strong interaction phenomena make prediction and explanation impractical (impossible)?  For instance, is it pointless to attempt the QCD-connected prediction of any nucleon structural property on a domain that is not yet empirically accessible?

If a reductive explanation is impossible, then science must rely on a tower of EFTs, each level developed for a different energy domain, in~order, e.g.,\ to express and understand the consequences of the emergence of mass and contingent effects, such as confinement, without~identifying their~source.

On the other hand, if~a reductive approach is possible, then non-perturbative calculational methods must be developed to define and solve QCD.  Prominent amongst such techniques today are (\emph{i}) the numerical simulation of lQCD~\cite{Gattringer:2010zz, Philipsen:2012nu, Banuls:2019rao} and (\emph{ii}) continuum Schwinger function methods (CSMs), viz.\~a~collection of models and schemes, each with varying degrees of connection to Equation\,\eqref{QCDdefine}.  Currently, each of these two approaches has strengths and weaknesses, so the best way forward is to combine them to the fullest extent that is reasonably possible and exploit the synergies that~emerge.

Among CSMs, the~Dyson--Schwinger equations (DSEs) have proven useful~\cite{Roberts:1994dr, Roberts:2000aa, Maris:2003vk, Chang:2011vu, Roberts:2012sv, Roberts:2015lja, Horn:2016rip, Eichmann:2016yit, Burkert:2017djo, Fischer:2018sdj, Qin:2020rad}.  These~quantum field theory generalisations of the Euler--Lagrange equations provide a continuum method for calculating Schwinger functions; namely, the~same Euclidean-space Green functions that are computed using lQCD.  Consequently, there are many opportunities for cross-fertilisation between DSE and lQCD studies, and this has been exploited to increasingly good effect during the past twenty years, especially at the level of propagators ($2$-point functions) and vertices ($3$-point functions) defined using QCD's elementary~degrees-of-freedom.

The challenge to DSE analyses is found in the fact that the equation for any given $n$-point function is coupled with those for some higher-$n$-point functions, e.g.,\ the~gap equation for the quark $2$-point function is coupled to those for the gluon $2$-point function and the gluon-quark $3$-point function.  Therefore, truncations are necessary in order to define a tractable problem.  Systematic, symmetry-preserving schemes have been developed~\cite{Munczek:1994zz, Bender:1996bb, Maris:1997hd, Chang:2009zb, Qin:2014vya, Williams:2015cvx, Binosi:2016rxz}, so that today DSE predictions can be distributed into three classes:
(\emph{A}) model-independent results in QCD;
(\emph{B}) illustrations of such results, using well-constrained model elements and possessing a recognisable connection to QCD;
and (\emph{C}) analyses that can reasonably be described as QCD-based, but~whose elements have not been calculated using a truncation that maintains a systematically-improvable connection with QCD.
Regarding the last two classes, comparisons between schemes and orders within schemes can be used to identify robust outcomes.  Results can also be compared with lQCD predictions, when available, capitalising on the overlap domain of these two approaches, and, of~course, predictions can be tested against experiment, which~is, as~always, the~final arbiter in~physics.

\subsection{Gluon~Mass}
\label{secGluonMass}
It is now possible and appropriate to return to \emph{the confinement hypothesis}, introduced in the opening paragraph of Section\,\ref{secConfinement} and expressed in the ``Millennium Problem''.  Beginning with  pioneering efforts roughly forty years ago~\cite{Cornwall:1981zr, Gribov:1999ui}, continuum and lattice studies of QCD's gauge sector have been growing in sophistication and reliability.  The~current state of understanding can be traced from an array of sources, see, e.g.,\ in \,\cite{Cornwall:1981zr, Gribov:1999ui, Dudal:2003by, Bowman:2004jm, Luna:2005nz, Aguilar:2008xm, RodriguezQuintero:2010wy, Boucaud:2011ugS, Strauss:2012dg, Binosi:2014aea, Aguilar:2015bud, Siringo:2015wtx, Binosi:2016nme, Cyrol:2016tym, Gao:2017uox, Binosi:2019ecz, Rodriguez-Quintero:2018wma, Cui:2019dwvS}.
Of specific interest is the property that the gluon propagator saturates at infrared momenta; to wit,
\begin{equation}
\label{eqGluonMass}
\Delta(k^2\simeq 0) = 1/m_g^2,
\end{equation}
where $\Delta(k^2)$ is the scalar function that characterises the dressed gluon propagator.  This~entails that the long-range propagation characteristics of gluons are markedly altered by their self-interactions.  Significantly, one may associate a renormalisation-group-invariant (RGI) gluon mass-scale with this effect: $m_0 \approx m_p/2$, and~summarise a large body of work by recording that gluons, although~behaving as massless entities on the perturbative domain, actually possess a running mass, with~$m_0$ characterising its value at infrared~momenta.

Asymptotic freedom guarantees that QCD's ultraviolet behaviour is tractable; but the emergence of a gluon mass reveals a new SM physics frontier because the existence of a running gluon mass, sizeable at infrared momenta, influences all analyses of the continuum bound-state problem.  For~instance, it could be a harbinger of gluon saturation~\cite{Accardi:2012qut, Brodsky:2015aiaS}.

Furthermore, $m_0>0 $ entails that QCD dynamically generates its own infrared cut-off, so that gluons with wavelength $\lambda \gtrsim \sigma :=1/m_0 \approx 0.5\,$fm decouple from the strong interaction, hinting at a dynamical realisation of confinement~\cite{Munczek:1983dx, Cahill:1985mh, Stingl:1985hx, Krein:1990sf, Burden:1991gd, Hawes:1993ef, Maris:1994ux, Roberts:1994dr, Bhagwat:2002tx, Roberts:2007ji, Bashir:2009fv, Strauss:2012dg, Bashir:2013zha, Qin:2013ufa, Lowdon:2015fig, Lucha:2016vte, Binosi:2016xxu, Binosi:2019ecz}.
In this picture, once a gluon or quark is produced, it starts to propagate in spacetime, but following each ``step'' of average length $\sigma$ an interaction occurs and the parton loses its identity, sharing it with others.  Ultimately a parton cloud is produced, which~fuses into colour-singlet final states.  This~physics is embodied in parton fragmentation functions (PFFs), which~describe how QCD partons, (nearly) massless when produced in a high-energy event, transform into a cascade of massive hadrons.  PFFs express the emergence of hadrons with mass from massless partons~\cite{Field:1977fa}.  Such observations suggest that PFFs are the cleanest expression of dynamical confinement in QCD.  This~perspective can be explored at modern and anticipated~facilities.

\subsection{Effective~Charge}
\begin{figure}[H]
\centering
\includegraphics[width=0.5\linewidth]{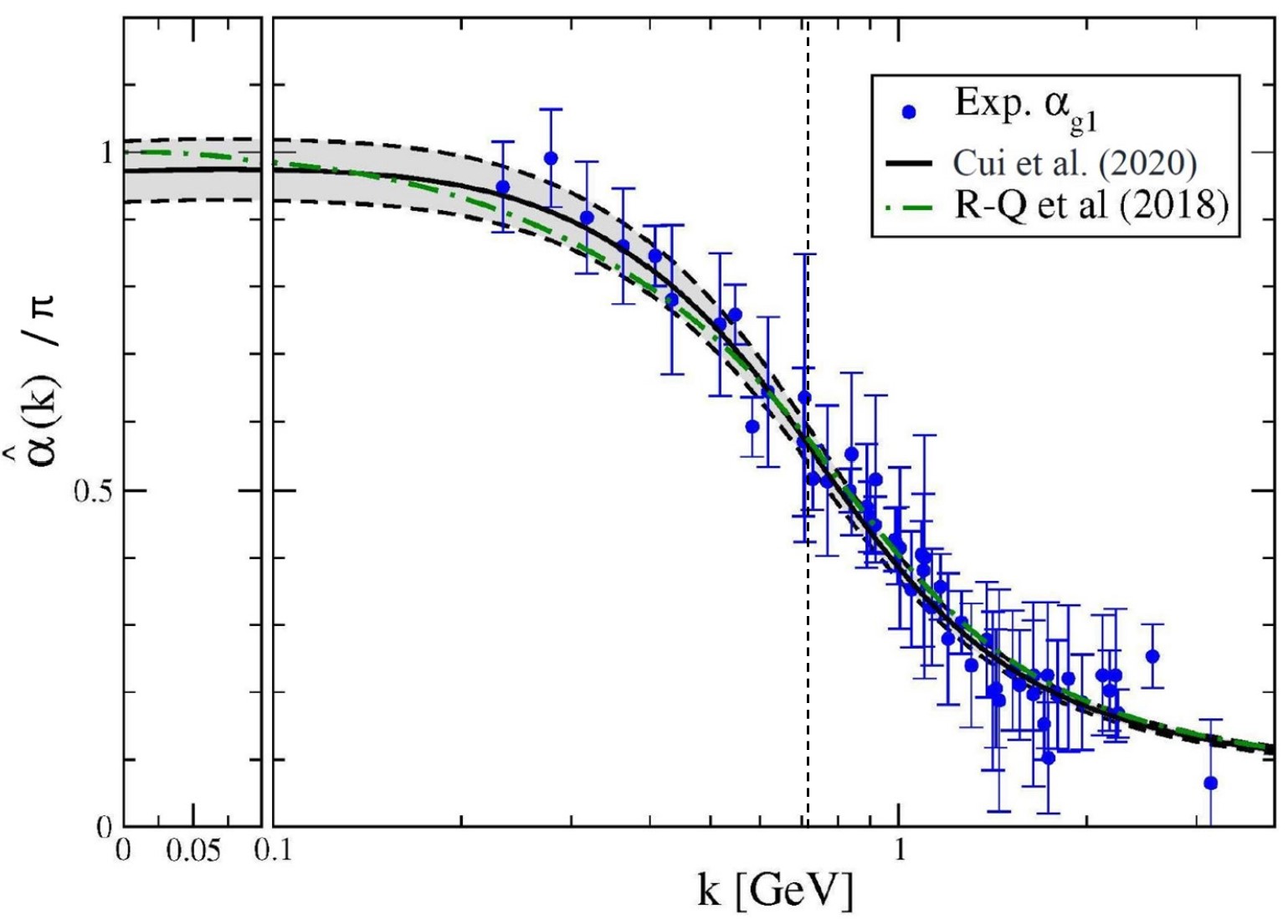}
\caption{
Solid black curve within grey band---RGI PI running-coupling, $\hat{\alpha}(k^2)/\pi$, computed in \,\cite{Cui:2019dwvS} (Cui~et~al.\ 2020), and dot-dashed green curve---earlier result (R-Q~et~al.\ 2018) \cite{Rodriguez-Quintero:2018wma}.
(The~grey band bordered by dashed curves indicates the uncertainty in the result arising from that in both continuum and lattice-QCD inputs and is detailed in \,\cite{Cui:2019dwvS}.)
For comparison, world data on the process-dependent charge, $\alpha_{g_1}$, defined via the Bjorken sum rule, are also depicted.  (The data sources are listed elsewhere~\cite{Cui:2019dwvS}.
For additional details, see e.g.,  in \,\cite{Deur:2005cf, Deur:2008rf, Deur:2016tte}.)
The k-axis scale is linear to the left of the vertical partition, and logarithmic otherwise.  The~vertical line,  $k=m_G$, marks the gauge sector screening mass, Equation\,\eqref{setzetaH}.
}
\label{Figwidehatalpha}
\end{figure}

Among the many other consequences of QCD's intricate non-perturbative gauge-sector dynamics is the generation of a process-independent (PI) running coupling, $\hat{\alpha}(k^2)$, see, e.g., in \,\cite{Binosi:2016nme, Rodriguez-Quintero:2018wma, Cui:2019dwvS}.  Depicted~as the solid (black) curve in Figure\,\ref{Figwidehatalpha}, this is a new type of effective charge.  It is an analogue of the Gell--Mann--Low effective coupling in QED~\cite{GellMann:1954fq} because it is completely determined by the gauge-boson vacuum polarisation, when the problem is approached using the pinch technique~\cite{Cornwall:1981zr, Cornwall:1989gv, Pilaftsis:1996fh, Binosi:2009qm} and background field method~\cite{Abbott:1980hw}.  The~result in Figure\,\ref{Figwidehatalpha} is a parameter-free Class-A DSE prediction, capitalising on analyses of QCD's gauge sector undertaken using continuum methods and informed by numerical simulations of~lQCD.

The data in Figure\,\ref{Figwidehatalpha} represent empirical information on $\alpha_{g_1}$, a~process-\emph{dependent} effective-charge~\cite{Grunberg:1982fw, Dokshitzer:1998nz} determined from the Bjorken sum rule, a~basic constraint on our knowledge of nucleon spin structure.  Solid theoretical reasons underpin the almost precise agreement between $\hat{\alpha}$ and $\alpha_{g_1}$ \cite{Binosi:2016nme, Rodriguez-Quintero:2018wma, Cui:2019dwvS},
%
so that the Bjorken sum serves as a window through which to gain empirical insight into QCD's effective~charge.

Figure\,\ref{Figwidehatalpha} shows that QCD's unique effective coupling is everywhere finite, i.e.,\ there is no Landau pole and the theory plausibly possesses an infrared-stable fixed point.  Apparently, QCD is infrared-finite because a gluon mass-scale is dynamically generated. %
(%
A theory is said to possess a Landau pole at $k^2_{\rm L}$ if the effective charge diverges at that point.  In~QCD perturbation theory, such a pole exists at $k^2_L=\Lambda_{\rm QCD}^2$.  Were such a pole to persist in a complete treatment of QCD, it would signal an infrared failure of the theory.  On~the other hand, the~absence of a Landau pole supports a view that QCD is alone amongst four-dimensional quantum field theories in being defined and internally consistent at all energy scales.  This~might have implications for attempts to develop an understanding of physics beyond the SM based upon non-Abelian gauge theories~\cite{Binosi:2016xxu, Appelquist:1996dq, Sannino:2009za, Appelquist:2009ka, Hayakawa:2010yn, Cheng:2013eu, Aoki:2013xza, DeGrand:2015zxa}).
In this case, the~value of the PI charge at $k^2=\Lambda_{\rm QCD}^2$ defines a screening mass~\cite{Cui:2020dlmM, Cui:2020piKM}: $m_G \approx 1.4 \,\Lambda_{\rm QCD} \approx 0.71\,$GeV.  As~evident in Figure\,\ref{Figwidehatalpha}, $m_G$ marks a boundary: the running coupling alters character at $k \simeq m_G$ so that modes with $k^2 \lesssim m_G^2$ are screened from interactions and the theory enters a practically conformal domain.  Evidently, the~line $k=m_G$ draws a natural border between soft and hard physics; therefore, defines the ``hadronic scale'':
\begin{equation}
\label{setzetaH}
\zeta_H=m_G\,.
\end{equation}

This is the scale at which all the properties of a hadron are expressed by the dressed quasiparticles that form the DSE kernels and emerge as the self-consistent solutions of the associated~equations.

As a unique PI effective charge, $\hat{\alpha}$ appears in each of QCD's dynamical equations of motion, including the gauge- and matter-sector gap equations, setting the strength of all interactions.  It~therefore plays a critical role in settling the fate of chiral symmetry; to wit, the~dynamical origin of light-quark masses in the SM even in the absence of a Higgs~coupling.

\subsection{Dynamical Chiral Symmetry~Breaking}
Just as a gluon mass-scale emerges dynamically in QCD, massless current-quarks become massive dressed-quarks through a phenomenon known as dynamical chiral symmetry breaking (DCSB) \cite{Nambu:2011zz}.  This~effect is another of the critical emergent phenomena in QCD.  It is expressed in hadron wave functions, not in vacuum condensates~\cite{Brodsky:2009zd, Brodsky:2010xf, Chang:2011mu, Brodsky:2012ku, Roberts:2015lja}, and modern theory indicates that more than 98\% of the visible mass in the Universe can be attributed to DCSB.  As classical massless-QCD is a scale-invariant theory (Section\,\ref{secWhence}),  this means that DCSB is fundamentally connected with the \emph{origin of mass from nothing}.

\begin{figure}[H]
\centerline{\includegraphics[clip,width=0.50\linewidth]{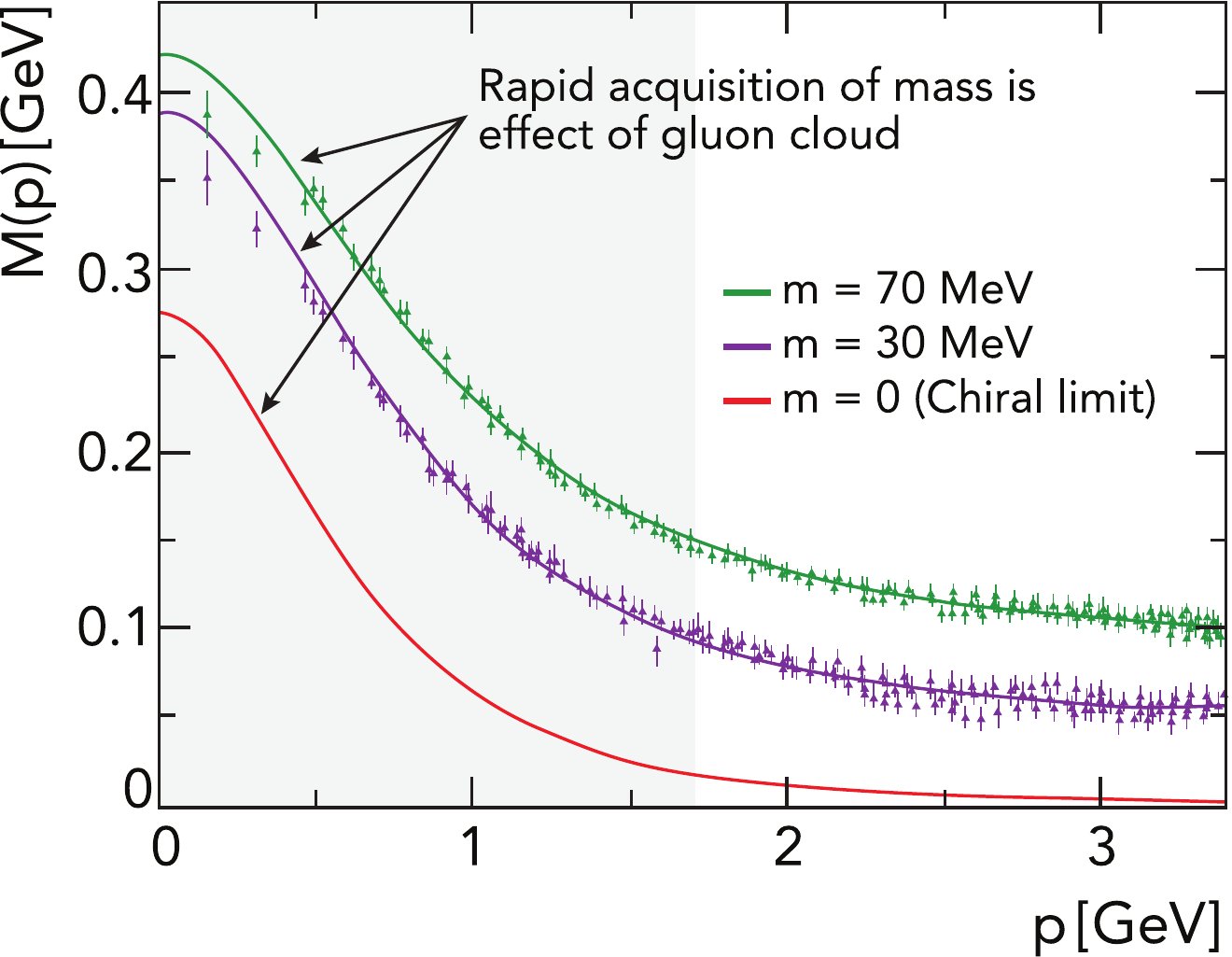}}
\caption{
Renormalisation-group-invariant dressed-quark mass function, $M(p)$ in Equation\,\eqref{Spgen}: \emph{solid~curves}---gap equation solutions~\cite{Bhagwat:2003vw, Bhagwat:2006tu}, ``data''---numerical simulations of lQCD \protect~\cite{Bowman:2005vx}, available~for current-quark masses $m=30,70$ MeV.
QCD's current-quark evolves into a constituent-quark as its momentum becomes smaller.  The~constituent-quark mass arises from a cloud of low-momentum gluons attaching themselves to the current-quark.  This~is DCSB, the~essentially non-perturbative effect that generates a quark \emph{mass} \emph{from nothing}; namely, it occurs even in the chiral limit.
Notably, the~size of $M(0)$ is a measure of the magnitude of the QCD scale anomaly in $n=1$-point Schwinger functions~\cite{Roberts:2016vyn}.
Moreover, experiments on $Q^2\in [0,12]\,$GeV$^2$ at the modern Thomas Jefferson National Accelerator Facility (JLab) will be sensitive to the momentum dependence of $M(p)$ within a domain that is here indicated approximately by the shaded region.
}
\label{gluoncloud}
\end{figure}

DCSB is most readily apparent in the dressed-quark propagator
\begin{equation}
\label{Spgen}
S(p) = 1/[i \gamma\cdot p A(p^2) + B(p^2)] = Z(p^2)/[i\gamma\cdot p + M(p^2)]\,,
\end{equation}
which is obtained as the solution of a gap equation whose kernel is critically dependent upon $\hat{\alpha}$.  $M(p^2)$ in Equation\,\eqref{Spgen} is the dressed-quark mass-function, depicted and explained in Figure\,\ref{gluoncloud}.

One must insist that chiral symmetry breaking in the absence of a Higgs mechanism is \underline{dynamical}.   It is distinct from Higgs-induced spontaneous symmetry breaking because (\emph{a}) nothing is added to QCD to catalyse this remarkable outcome and (\emph{b}) no simple change of variables in the QCD action can reveal it.  Instead, following quantisation of classical chromodynamics, with~its massless gluons and quarks, a~large mass-scale emerges in both the gauge- and~matter-sectors.

DCSB is empirically revealed very clearly in properties of the pion, whose structure in QCD is described by a Bethe--Salpeter amplitude:
\begin{equation}
\Gamma_{\pi}(k;P)  = \gamma_5 \left[
i E_{\pi}(k;P) + \gamma\cdot P F_{\pi}(k;P)  +\, \gamma\cdot k \, G_{\pi}(k;P) + \sigma_{\mu\nu} k_\mu P_\nu H_{\pi}(k;P) \right],
\label{genGpi}
\end{equation}
where $k$ is the relative momentum between the pion's valence-quark and -antiquark (defined such that the scalar functions in Equation\,\eqref{genGpi} are even under $k\cdot P \to - k\cdot P$) and $P$ is their total momentum.  $\Gamma_{\pi}(k;P)$ is directly related to an entity that, if~a nonrelativistic limit were appropriate, would become the pion's Schr\"odinger wave~function.

In chiral-limit QCD, if, and~only if, chiral symmetry is dynamically broken, then~\cite{Maris:1997hd, Qin:2014vya, Binosi:2016rxz}
\begin{equation}
\label{gtrE}
f_\pi^0 E_\pi(k;0) = B(k^2)\,,
\end{equation}
where $f_\pi^0$ is the pion's leptonic decay constant evaluated in the chiral-limit.
This identity is exceptional.
%
It is independent of the renormalisation scheme and true in any covariant gauge, and it entails that the two-body problem is solved, nearly completely, once the solution to the one body problem is known.  Furthermore, Equation\,\eqref{gtrE} has many corollaries, e.g.,\ it ensures that chiral-QCD generates a massless pion in the absence of a Higgs mechanism; predicts $m_\pi^2 \propto m$ on $m \simeq 0$, where $m$ is the current-quark mass; and entails that the chiral-limit leptonic decay constant vanishes for all excited-state pseudoscalar mesons with nonzero isospin~\cite{Holl:2004fr, Holl:2005vu}.  It is also the keystone that supports the success of chiral effective field theories in nuclear~physics.

Equation\,\eqref{gtrE}, which~may be described as a quark-level Goldberger--Treiman relation, is the most basic statement in QCD of the Nambu--Goldstone theorem~\cite{Nambu:1960tm, Goldstone:1961eq}. (Additional considerations applying to the $\eta$-$\eta^\prime$ complex are described elsewhere~\cite{Bhagwat:2007ha, Ding:2018xwy}.)
\begin{quote}
\emph{The Nambu--Goldstone theorem is fundamentally an expression of equivalence between the one-body problem and the two-body problem in QCD's colour-singlet  pseudoscalar channel.}
\end{quote}
It means that pion properties are practically a direct measure of the dressed-quark mass function rendered in Figure\,\ref{gluoncloud}.  Thus, the qualities of the nearly-massless pion are, enigmatically, the~cleanest expression of the mechanism that is responsible for almost all visible mass in the~Universe.

Reinstating the Higgs mechanism, so that the light-quarks possess their small current-masses, roughly commensurate with the electron mass, then DCSB is responsible for, inter alia, the physical size of the pion mass ($m_\pi \approx 0.15 \,m_p$); the large mass-splitting between the pion and its valence-quark spin-flip partner, the~$\rho$-meson ($m_\rho > 5 \,m_\pi$); and the natural scale of nuclear physics, $m_p \approx 1\,$GeV.  Curious things also happen to the kaon.  Similar to a pion, except~that a strange quark replaces one of the light quarks, the~kaon comes to possess a mass $m_K \approx 0.5\,$GeV.  In~this case, a~competition is taking place between dynamical and Higgs-driven mass generation~\cite{Ding:2015rkn, Chen:2016sno, Gao:2017mmp, Ding:2018xwy, Cui:2020dlmM, Cui:2020piKM}.

Expanding upon these observations, it is worth highlighting that the physical size of $m_\pi$ is actually much bigger than that linked with the Higgs mechanism for light quarks.  Empirically, the~scale of the Higgs effect in the light-quark sector is $\sim$1\,MeV~\cite{Zyla:2020}.  As~remarked above, $m_\pi = 0$ without a Higgs mechanism, but the current-masses of the light quarks in the pion are the same as they are in nucleons.  Therefore, the~simple Higgs mechanism result is $m_\pi \approx (m_u + m_d)$, yielding a value which is only 5\% of the physical mass.  This~physical mass emerges as the result of a hefty DCSB-induced enhancement factor, which~multiplies the current-quark mass contribution~\cite{Roberts:2020udq}.  However, the~scale of DCSB is $\sim m_p/3$, i.e.,\ the~size of a typical $u$ or $d$ constituent-quark mass, and the special NG-character of the pion means that although it \emph{should} have a mass $\sim (2/3) m_p \approx m_\rho$,  most of that mass is cancelled by gluon binding effects owing to constraints imposed by DCSB~\cite{Roberts:2016vyn}.  The~mechanism will now be~explained.

\subsection{Pion and the Trace~Anomaly}
\label{secPionTrace}
At this point, it is possible to resolve the dichotomy expressed by Equations\,\eqref{anomalyproton} and \eqref{EPTpion}.  These~statements hold equally on a measurable neighbourhood of the chiral limit because each hadron's mass is a continuous function of the current-masses of its valence-quarks/antiquarks.  So consider that for any meson, $H_2$, constituted from a valence-quark with current-mass $m_{q}$ and a valence-antiquark with mass $m_{\bar q}$,
\begin{equation}
\label{sigmaterm}
s_{H_2}(0) = \langle H_2(q) |  m_\Sigma \bar \psi \psi | H_2(q) \rangle = m_\Sigma  \frac{\partial m_{H_2}^2}{\partial m_\Sigma } \,,
\end{equation}
where $m_\Sigma=(m_q+m_{\bar q})$, viz.\ the scalar form factor at zero momentum transfer measures the reaction of the meson's mass-\emph{squared} to a variation in current-quark mass.  It is merely a misleading \emph{convention} to define the meson's $\sigma$-term as $\sigma_{H_2} = s_{H_2}(0)/[2 m_{H_2}]$.  Notably, the~pion (and any other NG mode) possesses the peculiar property that
\begin{equation}
s_\pi(0) \stackrel{m_\Sigma \simeq 0}{=} m_\Sigma \frac{\partial m_\pi^2 }{\partial m_\Sigma} = 1 \times m_\pi^2 ,
\end{equation}
which is the statement that in the vicinity of the chiral limit, 100\% of the pion mass-squared owes to the existence of the current-mass in ${\mathpzc L}_{\rm QCD}$.  One should compare this result with that for the pion's valence-quark spin-flip partner, i.e.,\ the~$\rho$-meson~\cite{Flambaum:2005kc}:
\begin{equation}
s_\rho(0)  \approx 0.06 \, m_\rho^2,
\end{equation}
indicating that just 6\% of the $\rho$-meson's mass-squared is generated by the current-mass term in ${\mathpzc L}_{QCD}$.  The~remainder arises largely as a consequence of EHM~\cite{Roberts:2020udq}, as~suggested by the fact that $m_\rho \approx 2 M(0)$, where $M(k^2)$ is the dressed-quark mass function in Figure\,\ref{gluoncloud}.  The~key to understanding Equation\,\eqref{EPTpion} is Equation\,\eqref{gtrE} and three associated Gold\-berger--Treiman-like relations, which~are exact in chiral QCD.  Utilising these identities when working with those DSEs needed to describe a pseudoscalar meson, an~algebraic proof of the following statement can be constructed~\cite{Maris:1997hd, Qin:2014vya, Binosi:2016rxz}: at every order in a symmetry-preserving analysis, the~masses generated for the valence-quarks that constitute the system are exactly cancelled by the attraction produced by interactions between them.  This~cancellation ensures that the initial two-valence-parton system, which~began massless, grows~into a complex system, with~a bound-state wave function tied to a pole in the scattering matrix at $P^2=0$.  Stated~simply,  Equation\,\eqref{EPTpion} is obtained through cancellations between one-body dressing and two-body binding~effects:
\begin{equation}
M^{\rm dressed}_{\rm quark} + M^{\rm dressed}_{\rm antiquark} + U^{\rm dressed}_{\rm quark-antiquark\;interaction} \stackrel{\rm chiral\;limit}{\equiv} 0\,,
\label{EasyOne}
\end{equation}
with the sum being exactly zero if, and~only if, chiral symmetry is broken dynamically in the Nambu pattern. (A full discussion is provided in Sec.\,3.3 of Ref.\,\cite{Roberts:2016vyn}.) 
Away from the chiral limit for NG modes and always in other channels, the~cancellation is~incomplete.

It is important to remark that such a transparent resolution of the conundrum expressed by Equations\,\eqref{anomalyproton} and \eqref{EPTpion} is impossible if one insists on using a parton model basis, in~which the trace anomaly operator is given simply by Equation\,\eqref{SIQCD}.  One must employ a modern quasiparticle formalism.  Then, with dressed-particle propagators and bound-state wave functions, obtained at a low renormalisation scale, $\zeta \lesssim m_p$, as~solutions of coupled integral equations, each of which sums a countable infinity of diagrams, Equation\,\eqref{EPTpion}
can be re-expressed:
\begin{subequations}
\label{QuasiParticle}
\begin{align}
\langle \pi(q) & | \Theta_0 | \pi(q) \rangle
 \stackrel{\zeta \gg m_p}{=} \langle \pi(q) | \tfrac{1}{4} \beta(\alpha(\zeta)) G^{a}_{\mu\nu}G^{a}_{\mu\nu} | \pi(q) \rangle  \stackrel{\zeta \lesssim m_p}{\to}
\langle \pi(q) | {\cal D}_1 + {\cal I}_2 |\pi(q)\rangle\,, \\
{\cal D}_1 & = \sum_{f=u,d} M_f(\zeta) \, \bar {\cal Q}_f(\zeta) {\cal Q}_f(\zeta) \,, \quad
{\cal I}_2  = \tfrac{1}{4} [\beta(\alpha(\zeta)) {\cal G}^{a}_{\mu\nu}{\cal G}^{a}_{\mu\nu}]_{2{\rm PI}}  \,.
\end{align}
\end{subequations}

Equations~\eqref{QuasiParticle} describe the transfigurement of the chiral-limit parton-basis expression of the trace anomaly's expectation value in the pion into a new form, written in terms of a non-perturbatively-dressed quasi-particle basis, with~${\cal Q}$ denoting dressed-quarks and ${\cal G}$ the dressed-gluon field strength tensor.
Here, the~first term expresses the one-body-dressing content of the trace anomaly.  It is positive.  Patently, a~massless valence-quark (antiquark) gaining a large mass by way of interactions with its own gluon field is an effect of the trace-anomaly in what may be described as the one-quasiparticle subspace of a complete pion wave function.
The second term expresses the two-particle-irreducible (2PI) interaction content of the forward scattering process represented by this trace anomaly matrix element.  It is negative and acquires a scale because the gluon- and quark-propagators and the couplings in the 2PI processes have all gained a~mass-scale.

The discussion of Equations\,\eqref{EasyOne} and \eqref{QuasiParticle} and their connection with Equation\,\eqref{EPTpion} bring a well-known adage to mind, quoted here from S.~Weinberg in \,\cite{Guth:1984rq}: ``You may use any degrees of freedom you like to describe a physical system, but~if you use the wrong ones, you’ll be sorry!''

\section{Empirical Manifestations of Emergent~Mass}
\unskip
\subsection{Pion Wave~Function}
\label{secPDA}
Empirical signals of the emergence of mass are ubiquitous, but~their appearance typically changes from one system to another.  This~means that it is essential to study a wide variety of observables because each one is apt to expose different aspects of the underlying mechanisms.  It is useful, therefore, to~begin with the most obvious, whose importance derives from Equation\,\eqref{gtrE}, i.e.,\ the~pion's leading-twist two-particle parton distribution amplitude (PDA), $\varphi_\pi$, the~simplest component of its light-front wave function.  Any framework that is capable of delivering a hadron's Poincar\'e-covariant bound-state amplitude can also be used to calculate its PDA, and in this case, $\varphi_\pi$ is given by a light-front projection of the pion's Bethe--Salpeter wave function~\cite{Chang:2013pqS}
\begin{equation}
f_\pi\, \varphi_\pi(x;\zeta)  = {\rm tr}_{\rm CD}
Z_2 \! \int_{dq}^\Lambda \!\!
\delta(n\cdot q_+ - x \,n\cdot P) \, \gamma_5\gamma\cdot n\,
S(q_+)\Gamma(q;P)S(q_{-})\,.
\label{pionPDA}
\end{equation}

In Equation\,\eqref{pionPDA}, the trace is over colour and spinor indices; $\int_{dq}^\Lambda$ is a Poincar\'e-invariant regularisation of the four-dimensional integral, with~$\Lambda$ the ultraviolet regularization mass-scale; $Z_{2}(\zeta,\Lambda)$ is the quark wavefunction renormalisation constant; $n$ is a light-like four-vector, $n^2=0$; and $P$ is the pion's four-momentum, $P^2=-m_\pi^2$ and $n\cdot P = -m_\pi$ in the pion's rest~frame.

\begin{figure}[t]
\centering
\includegraphics[width=0.6\textwidth,clip]{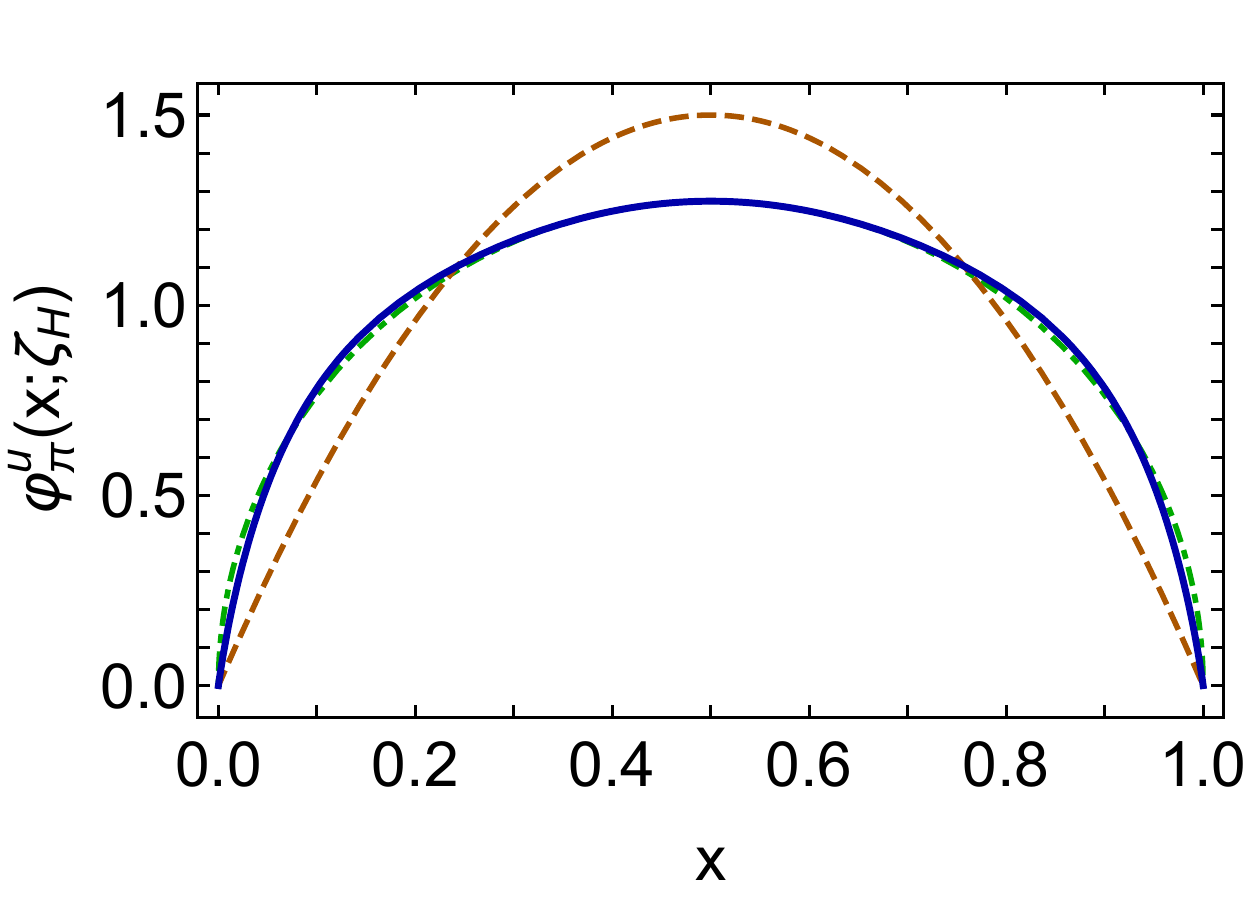}
\caption{
Twist-two pion PDA at the hadronic scale, $\zeta_H$.
Solid blue curve, Equation\,\eqref{pionDADB}---determined using the most sophisticated available symmetry preserving DSE kernels;
dot-dashed green curve---original prediction from the work in \,\cite{Chang:2013pqS}.%
~These PDA results are consistent with contemporary lQCD results~\cite{Segovia:2013ecaS, Shi:2014uwaS}.
The dashed orange curve is $\varphi^{\rm asy}_\pi(x)=6 x(1-x)$, the~limiting form under QCD evolution~\cite{Lepage:1979zb, Efremov:1979qk, Lepage:1980fj}.
The PDAs are dimensionless.}
\label{FigpionPDA}
\end{figure}

Two distinctively different truncations of QCD's DSEs~\cite{Chang:2013pqS} have been used to calculate the amplitude in Equation\,\eqref{pionPDA}.  Both agree: compared with the asymptotic profile, which~is valid on $\Lambda_{\rm QCD}/\zeta \simeq 0$, there is a marked broadening of $\varphi_\pi(x;\zeta)$, which~owes exclusively to DCSB, i.e.,\ the~emergence of mass as exhibited in Figure\,\ref{gluoncloud}.  This~causal connection may be asserted because the PDA is calculated at a low renormalisation scale in the chiral limit, in~which case the quark mass function owes entirely to DCSB via Equation\,\eqref{gtrE}.  Moreover, the~PDA's dilation is related to the rate at which a dressed-quark approaches the asymptotic bare-parton limit.  The~prediction determined using the most sophisticated kernel~\cite{Cui:2020dlmM, Cui:2020piKM} is depicted in Figure\,\ref{FigpionPDA}:
\begin{equation}
\varphi_\pi (x;\zeta_H) = 20.227\, x(1-x)
 [1- 2.5088\, \sqrt{x(1-x)}+ 2.0250 \, x(1-x)]\,,\label{pionDADB}
\end{equation}

It can be verified empirically at JLab\,12, e.g.,\ in measurements of the pion's electromagnetic form factor  (see Section\,\ref{secPionFF} below).

It is worth remarking here that the authors of \,\cite{Chang:2013pqS} chose to reconstruct the pion's DA from its Mellin moments using an order-$\alpha$ Gegenbauer expansion.  This~is useful as a first step because the procedure converges rapidly, so enables the qualitative feature of broadening driven by EHM in the SM to be exposed.  Such broadening, however, need not and should not disturb the DA's endpoint behaviour, which~QCD predicts to be linear in the neighbourhoods $x\simeq 0, 1$.~Therefore, as~a second step, \mbox{the authors of \,\cite{Cui:2020dlmM, Cui:2020piKM} re-expressed the result from the work in \,\cite{Chang:2013pqS}} as the function in Equation\,\eqref{pionDADB}.  The~first eleven moments agree at the level of 1.6 (1.4)\%, i.e.,\ well within any sensible estimate of uncertainty in the computation of high-order moments.  Pointwise, the~curves are practically indistinguishable, as~highlighted in Figure\,\ref{FigpionPDA}.

It is important to emphasise that the computed PDAs in Figure\,\ref{FigpionPDA} are concave functions.  Such~pointwise behaviour contrasts markedly with the ``bimodal'' or ``double-humped'' distributions that have been favoured in phenomenological applications by some authors~\cite{Chernyak:1983ej}.~It should be understood in this connection that a double-humped profile for the twist-two PDA places it in the class of distributions produced by a Bethe--Salpeter wave function which is suppressed at zero relative momentum, instead of maximal thereat.  No ground-state solution of the pseudoscalar or vector meson Bethe--Salpeter equation exhibits corresponding behaviour~\cite{Maris:1997tm, Maris:1999nt, Bhagwat:2006pu, Krassnigg:2009zh, Qin:2011xq}.  Thus, a~bimodal distribution cannot be an accurate pointwise rendering of the PDA for a ground-state meson.
Notwithstanding that, if~one employs such a distribution in a phenomenological application for which only its lowest Mellin moments are important,
then some carefully-constrained bimodal distributions may supply useful approximations to the moments of a broad, concave PDA and thereby prove practically~useful.

A question of more than thirty-years standing can be answered using Figure\,\ref{FigpionPDA}, namely, when~does $\varphi^{\rm asy}_\pi(x)$ provide a good approximation to the pion PDA?  Plainly, not at $\zeta_H$.  The~ERBL evolution equation~\cite{Lepage:1979zb, Efremov:1979qk, Lepage:1980fj} describes the $\zeta$-evolution of $\varphi_\pi(x;\zeta)$, and applied to $\varphi_\pi(x;\zeta)$ in Figure\,\ref{FigpionPDA}, one~finds~\cite{Segovia:2013ecaS, Shi:2014uwaS} that $\varphi^{\rm asy}_\pi(x)$ is a poor approximation to the true result even at $\zeta=10\,$GeV.  Thus~at energy scales accessible to experiment, the~twist-two PDAs of ground-state hadrons are ``fat and squat''.  Evidence supporting this picture had long been accumulating~\cite{Mikhailov:1986be, Petrov:1998kg, Braun:2006dg, Brodsky:2006uqa}, and the dilation is confirmed by simulations of lQCD~\cite{Segovia:2013ecaS, Shi:2014uwaS}.

\subsection{Pion Electromagnetic Form~Factor}
\label{secPionFF}
The cross sections for many hard exclusive hadronic reactions, i.e.,\ processes involving a highly energetic probe that strikes a target and leaves it intact, can be written in terms of the PDAs of the hadrons involved.  An~example is the elastic electromagnetic form factor of the pion, for~which the prediction can be stated thus~\cite{Efremov:1979qk, Lepage:1979zb, Lepage:1980fj}:
\begin{subequations}
\label{pionUV}
\begin{align}
\exists Q_0>\Lambda_{\rm QCD} \; & |\;  Q^2 F_\pi(Q^2) \stackrel{Q^2 > Q_0^2}{\approx} 16 \pi \alpha(Q^2)  f_\pi^2 \mathpzc{w}_\varphi^2,\\
\mathpzc{w}_\varphi & = \frac{1}{3} \int_0^1 dx\, \frac{1}{x} \varphi_\pi(x)\,,
\end{align}
\end{subequations}
where $\alpha(Q^2) $ is the running coupling, which~is practically equivalent to $\hat{\alpha}(Q^2)$ in {Figure}\,\ref{Figwidehatalpha}
on any domain within which perturbation theory is valid.  The~value of $Q_0$ is not predicted by perturbative QCD (pQCD), but it is computable in any non-perturbative framework that veraciously expresses~EHM.

It was anticipated that JLab would verify the fundamental prediction in Equation\,\eqref{pionUV}.  Indeed, the~first publication by the $F_\pi$ Collaboration~\cite{Volmer:2000ek} heralded the beginning of a new era in mapping the pion's internal structure.  Ensuing measurements~\cite{Horn:2006tm, Tadevosyan:2007yd, Horn:2007ug, Huber:2008id, Blok:2008jy} confirmed the data's trend, leading~to a common perception that, at~$Q^2=2.45\,$GeV$^2$, one remains far from the resolution field wherein the pion acts like an elementary partonic quark--antiquark pair, i.e.,\ far from validating Equation\,\eqref{pionUV}.  This~conclusion was based on the assumption that inserting $\varphi^{\rm asy}_\pi(x)$ into Equation\,\eqref{pionUV} delivers a valid approximation at $\zeta^2=Q^2=2.45\,$GeV$^2$, so that
\begin{equation}
\label{pionUV4}
Q^2 F_\pi(Q^2) \stackrel{Q^2=4\,{\rm GeV}^2}{\approx} 0.15\,.
\end{equation}

The result in Equation \,\eqref{pionUV4} is a factor of $2.7$ smaller than the empirical value quoted at $Q^2 =2.45\,$GeV$^2$ \cite{Huber:2008id}: $0.41^{+0.04}_{-0.03}$, and a factor of three smaller than that computed at $Q^2 =4\,$GeV$^2$ in \,\cite{Maris:2000sk}.  At~the time, the authors of \,\cite{Maris:2000sk} supplied the lone prediction for the $Q^2$-dependence of $F_\pi$ that was both applicable on the full spacelike region mapped reliably by experiment and confirmed~thereby.

Here, the~perception of a mismatch and a real discrepancy are not equivalent because, as~elucidated above, $Q^2=4\,$GeV$^2$ is not within the domain $\Lambda_{\rm QCD}^2/Q^2\simeq 0$ upon which Equation\,\eqref{pionUV} used with $\varphi^{\rm asy}_\pi(x)$ provides a valid approximation.  This~being so, and~given the successful prediction in \,\cite{Maris:2000sk}, it was natural to ask whether DSE methods could address the issue of the ultimate validity of Equation\,\eqref{pionUV}.

\begin{figure}[t]
\centerline{\includegraphics[clip,width=0.6\textwidth]{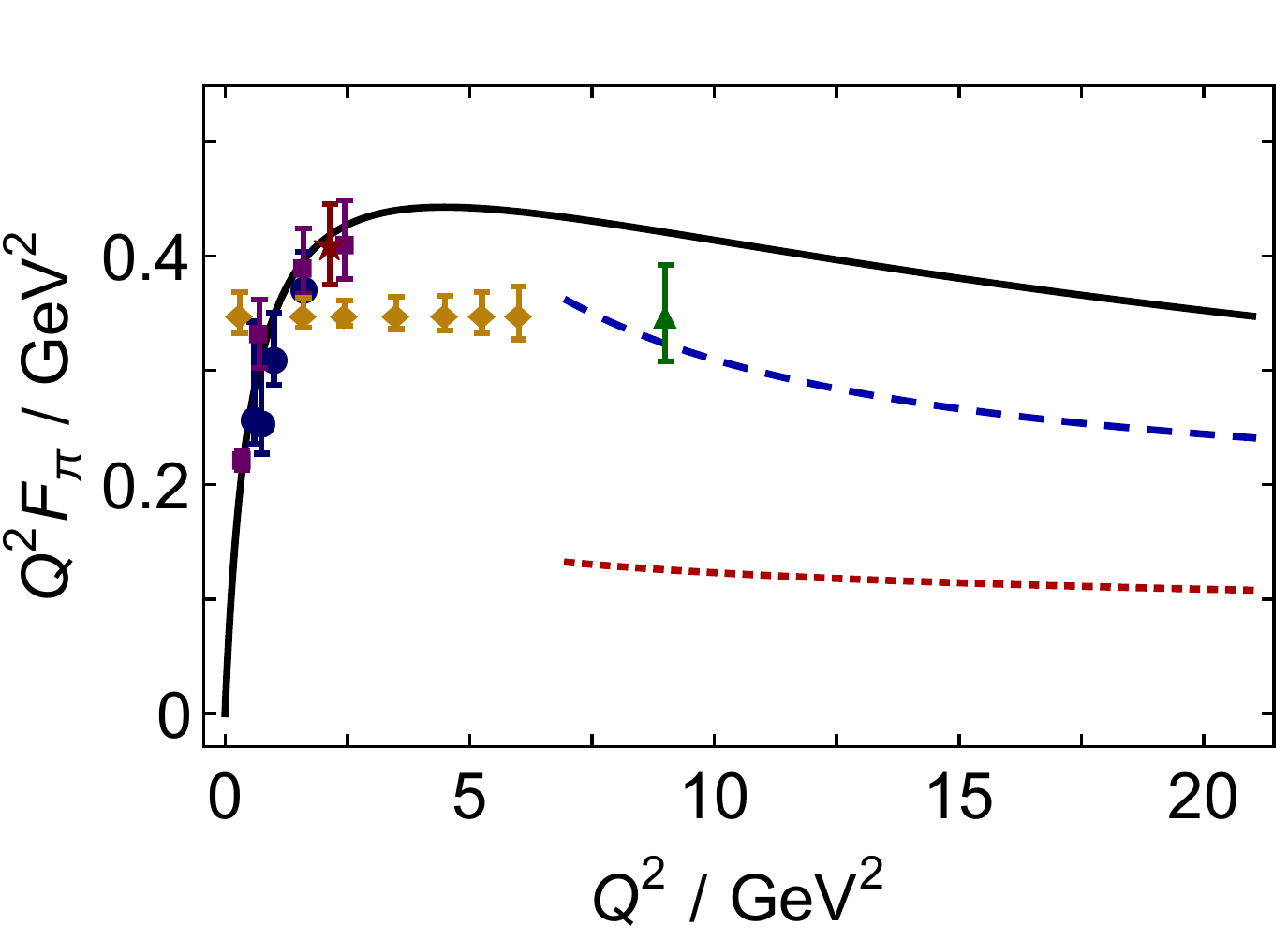}}
\caption{
$Q^2 F_\pi(Q^2)$.~Solid curve (black)---theoretical prediction~\cite{Chang:2013nia, Gao:2017mmp, Chen:2018rwz}; dashed curve (blue)---pQCD prediction computed with the modern, dilated pion PDA described in Section\,\protect\ref{secPDA}; and dotted (red) curve---pQCD prediction computed with the asymptotic profile, $\varphi^{\rm asy}_\pi(x)$, which~had previously been used to guide expectations for the asymptotic behaviour of $Q^2 F_\pi(Q^2)$.  The~filled-circles and \mbox{-squares} represent existing JLab data~\cite{Huber:2008id} and the filled diamonds and triangle, whose normalisations are arbitrary, indicate the projected $Q^2$-reach and accuracy of forthcoming experiments~\cite{E1206101, E1207105}.}
\label{figWPFpi}
\end{figure}

Developments within the past decade have made that possible and now a computation of the pion's electromagnetic form factor is available to arbitrarily large-$Q^2$ \cite{Chang:2013nia, Gao:2017mmp, Chen:2018rwz}.  The~result is the solid (black) curve in Figure\,\ref{figWPFpi}.
In addition, the~study enables that result to be correlated with Equation\,\eqref{pionUV} via the modern PDA computed in the same framework, which~is the dashed (blue) curve in Figure\,\ref{figWPFpi}.  On~the depicted domain, this leading-order, leading-twist QCD prediction, computed with a pion valence-quark PDA evaluated at a scale fitting the experiment, underestimates the full DSE result by a fairly uniform 15\%.  The~mismatch is explained by a mix of higher-order, higher-twist corrections to Equation\,\eqref{pionUV} in pQCD on the one hand and, on~the other hand, shortcomings in the leading-order DSE truncation used in \,\cite{Chang:2013nia, Gao:2017mmp, Chen:2018rwz}, which~predicts the right power-law behaviour for the form factor but not precisely the correct anomalous dimension (logarithm exponent) in the strong-coupling calculation.
It is now anticipated that the upgraded JLab facility will reveal a maximum in $Q^2 F_\pi(Q^2)$ at $Q^2 \approx 6\,$GeV$^2$ and an experiment at $Q^2=9\,$GeV$^2$ will see an indication of parton model scaling and scaling violations for the first time in a hadron elastic form factor.  While JLab's grip on these things might be tenuous, the~reach of an electron ion collider would certainly enable validation of these predictions~\cite{Aguilar:2019teb}.

\subsection{Valence-Quark Distributions in the~Pion}
Given the pions' simple valence-quark content, i.e.,\ one quark and one antiquark, a~basic quantity in any discussion of their structure is the associated parton distribution function (PDF), ${\mathpzc q}^\pi(x;\zeta)$.  This~density charts the probability that a valence ${\mathpzc q}$-quark in a pion carries a light-front fraction $x$ of the system's total momentum~\cite{Ellis:1991qj}, and one of the earliest predictions of pQCD is~\cite{Ezawa:1974wm, Farrar:1975yb, Berger:1979du}:
\begin{equation}
\label{PDFQCD}
{\mathpzc q}^{\pi}(x;\zeta =\zeta_H) \sim (1-x)^{2}\,.
\end{equation}

Moreover, the~exponent evolves as $\zeta$ increases beyond $\zeta_H$, becoming $2+\gamma$, where $\gamma\gtrsim 0$ is an anomalous dimension that increases logarithmically with $\zeta$ \cite{Gribov:1972, Lipatov:1974qm, Altarelli:1977, Dokshitzer:1977}.  (In the limit of exact ${\mathpzc G}$-parity symmetry, which~is a good approximation in the SM, $u^{\pi^+}(x) = \bar d^{\pi^+}(x)$, etc. Therefore, it is only necessary to discuss one unique distribution.)

As described in connection with Equations\,\eqref{EasyOne} and \eqref{QuasiParticle}, for~nature's pion there is near-complete cancellation between (\emph{a}) strong-mass-generating dressing of the valence-quark and -antiquark and (\emph{b}) binding attraction.  Such effects distinguish this system from the more massive kaon, within~which the cancellation is far less efficient because the $s$-quark current mass generated by the Higgs is so much larger than that of the $u$- and $d$-quarks.  Consequently, high-precision measurements of the valence-parton distributions in the pion and kaon are a high priority at existing and anticipated facilities~\cite{Keppel:2015, C12-15-006A, Denisov:2018unjF, Aguilar:2019teb}.

Such efforts are driven by ongoing progress in theory.%
~Marking one significant class of advances, lQCD is beginning to yield results for the pointwise behaviour of the pion's valence-quark distribution~\cite{Xu:2018eii, Chen:2018fwa, Karthik:2018wmj, Sufian:2019bol,  Izubuchi:2019lyk}, offering promise for information beyond the lowest few moments~\cite{Oehm:2018jvm, Joo:2019bzr}.

Extensions of the continuum analysis in \,\cite{Hecht:2000xa} are also yielding new insights~\cite{Chang:2014lvaS}, leading to the first parameter-free predictions of the valence, glue and sea distributions within the pion~\cite{Ding:2019qlr, Ding:2019lwe}, unifying them with, inter alia, electromagnetic pion elastic and transition form \mbox{factors~\cite{Chang:2013nia, Raya:2015gvaS, Raya:2016yuj, Gao:2017mmp, Chen:2018rwz, Ding:2018xwy}}.  The~analysis reveals that, like the pion's PDA in Figure\,\ref{FigpionPDA}, the~valence-quark distribution function is hardened by DCSB, producing the following apportioning of light-front momentum at the scale $\zeta=\zeta_2=2\,$GeV:
\begin{equation}
\label{momfractions}
\langle x_{\rm valence} \rangle = 0.48(3)\,,\quad
\langle x_{\rm glue} \rangle  = 0.41(2)\,,\quad
\langle x_{\rm sea} \rangle  = 0.11(2)\,.
\end{equation}

A similar valence-quark momentum fraction was obtained in Ref.\,\cite{Barry:2018ort} by analysing data on $\pi$-nucleus Drell--Yan and leading neutron electroproduction~\cite{Barry:2018ort}:  $\langle 2 x \rangle_{\mathpzc q}^\pi  = 0.49(1)$ at $\zeta=\zeta_2$.  Even~though this phenomenological PDF yields a compatible momentum fraction, its $x$-profile is different.  In~fact, the~phenomenological PDF conflicts with the QCD constraint, Equation\,\eqref{PDFQCD}.  Significantly, the~analysis in \,\cite{Barry:2018ort} ignored threshold resummation effects, which~are known to have a material impact at large $x$~\cite{Aicher:2010cb,Westmark:2017uig}.  (Similar remarks apply to the analysis in  \,\cite{Novikov:2020snp}.)

\begin{figure}[H]
\centerline{%
\includegraphics[clip, width=0.60\textwidth]{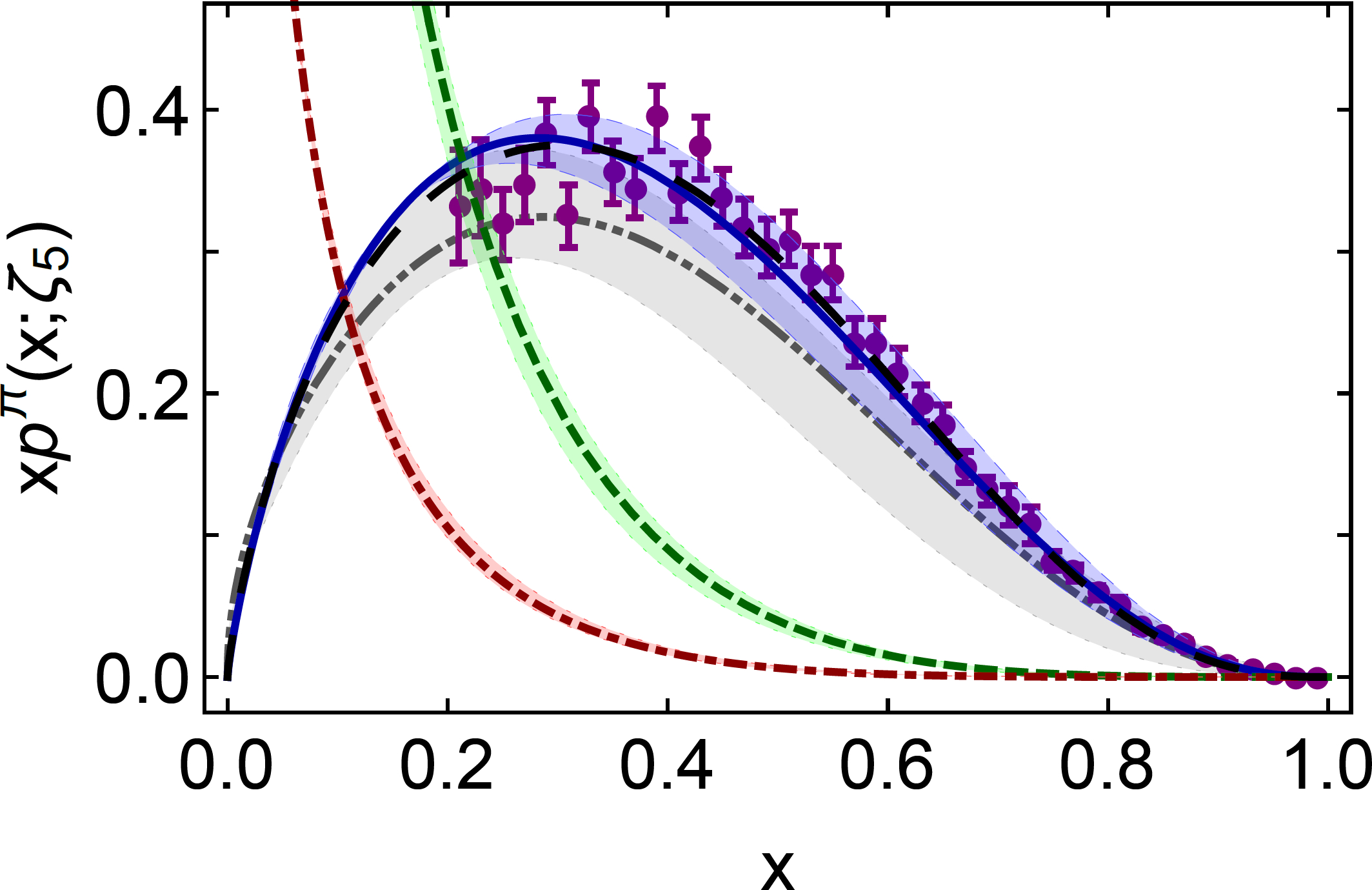}}
\caption{
Pion valence-quark momentum distribution function, $x {\mathpzc u}^\pi(x;\zeta_5)$:
dot-dot-dashed~(grey) curve within shaded band---lQCD result~\cite{Sufian:2019bol};
long-dashed (black) curve---early continuum analysis~\cite{Hecht:2000xa};
and solid (blue) curve within shaded band---modern, continuum calculation~\cite{Ding:2019qlr, Ding:2019lwe}.
From~\mbox{the works in \,\cite{Ding:2019qlr, Ding:2019lwe}}: gluon momentum distribution in pion, $x g^\pi(x;\zeta_5)$---dashed (green) curve within shaded band;
and sea-quark momentum distribution, $x S^\pi(x;\zeta_5)$---dot-dashed (red) curve within shaded band.
(The shaded bands indicate the size of calculation-specific uncertainties, detailed in the source \mbox{material~\cite{Sufian:2019bol, Ding:2019qlr, Ding:2019lwe}}.)
Data (purple) from in \,\cite{Conway:1989fs}, rescaled according to the analysis in \,\cite{Aicher:2010cb}.}
\label{figF12}
\end{figure}

Importantly, as~illustrated in Figure\,\ref{figF12}, after~evolution~\cite{Gribov:1972, Lipatov:1974qm, Altarelli:1977, Dokshitzer:1977} to $\zeta=\zeta_5=5.2\,$GeV, using $\hat{\alpha}(k^2)$ in Figure\,\ref{Figwidehatalpha} to integrate the evolution equations, the~continuum prediction for $u^\pi(x)$ from the works in \,\cite{Ding:2019qlr, Ding:2019lwe} matches that obtained using lQCD~\cite{Sufian:2019bol}. Given that no parameters were varied in order to procure this or any other outcome in \,\cite{Ding:2019qlr, Ding:2019lwe}, a~remarkable, modern confluence has been reached, which~suggests that real strides are being made toward understanding pion structure and its relation to the emergence of mass.  (Predictions for kaon PDFs are described in \,\cite{Cui:2020dlmM, Cui:2020piKM}.)

\subsection{Emergence of Diquark~Correlations}
The emergence of mass is also expressed in the properties of baryons, something which can readily be seen when the three valence-quark bound-state problem is studied with the same level of sophistication that is now typical for mesons.  In~this connection, DCSB, as~displayed in the momentum-dependence of the dressed-quark mass---Figure\,\ref{gluoncloud}, is as important to baryons as it is to mesons.  In~fact, one important consequence of DCSB is that any interaction able to create composite pseudo--NG modes from a light dressed-quark and -antiquark, and~reproduce the observed value of their leptonic decay constants, will also generate tight colour-antitriplet correlations between any two of a nucleon's dressed-quarks.
This conclusion is based upon evidence gathered in thirty years of studying two- and three-body bound-states in hadron physics~\cite{Barabanov:2020jvnS}.

The properties of such diquark correlations are known.  As~colour-carrying correlations, diquarks are confined~\cite{Bender:1996bb, Bhagwat:2004hn}.~Additionally, owing to properties of charge-conjugation, a~diquark with spin-parity $J^P$ may be viewed as a partner to the analogous $J^{-P}$ meson~\cite{Cahill:1987qr}.  It follows that the strongest diquark correlations in the nucleon are scalar isospin-zero, $[ud]_{0^+}$, and pseudovector, isospin-one, $\{uu\}_{1^+}$, $\{ud\}_{1^+}$, $\{dd\}_{1^+}$.  Furthermore, although~no pole-mass exists, the~following mass-scales, expressing the range and strength of the correlation, can be associated with these diquarks~\cite{Cahill:1987qr, Maris:2002yu, Eichmann:2008ef, Segovia:2015hraS, Segovia:2015ufa, Eichmann:2016hgl, Lu:2017cln, Chen:2017pse} (in GeV),
\begin{equation}
m_{[ud]_{0^+}} \approx 0.7-0.8\,,\;
m_{\{uu\}_{1^+}}  \approx 0.9-1.1  \,,
\end{equation}
with $m_{\{dd\}_{1^+}}=m_{\{ud\}_{1^+}} = m_{\{uu\}_{1^+}}$ in the isospin symmetric limit.  Notably, the~nucleon contains both scalar-isoscalar and pseudo\-vec\-tor-isovector correlations.  Neither can safely be ignored and their presence has many observable consequences~\cite{Roberts:2013mja, Segovia:2013ugaS, Segovia:2016zyc, Mezrag:2017znp}.  

Realistic diquark correlations are also soft and interacting.  They all carry charge, scatter~electrons, and~have an electromagnetic size which is similar to that of the kindred mesonic system, \mbox{see, e.g., in\ \cite{Maris:2004bp, Eichmann:2008ef, Roberts:2011wy}}:
\begin{equation}
\label{qqradii}
r_{[ud]_{0^+}} \gtrsim r_\pi, \quad r_{\{uu\}_{1^+}} \gtrsim r_\rho,
\end{equation}
with $r_{\{uu\}_{1^+}} > r_{[ud]_{0^+}}$.  As~in the meson sector, these scales are set by that associated with~DCSB.

It should be emphasised that such fully dynamical diquark correlations are vastly different from the static, pointlike ``diquarks'' introduced originally~\cite{Anselmino:1992vg} in an attempt to solve the so-called ``missing~resonance'' problem~\cite{Aznauryan:2011ub}, viz.\ the fact that quark models predict a far greater number of baryon states than were observed in the previous millennium~\cite{Burkert:2004sk}. Moreover, their existence enforces certain distinct interaction patterns for the singly- and doubly-represented valence-quarks within the proton, as~exhibited elsewhere~\cite{Roberts:2015lja, Roberts:2013mja, Segovia:2013ugaS, Segovia:2014aza, Segovia:2016zyc, Burkert:2017djo, Chen:2017pse, Chen:2018nsg, Chen:2019fzn, Lu:2019bjs, Cui:2020rmuS}.

\begin{figure}[H]
\centerline{%
\includegraphics[clip,width=0.60\textwidth]{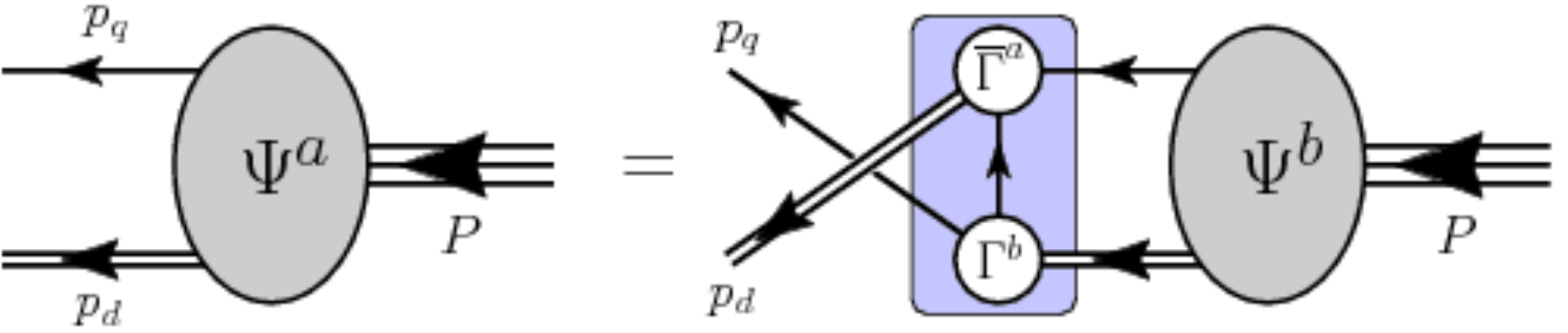}}
\caption{
Poincar\'e covariant Faddeev equation: a homogeneous linear integral equation for $\Psi$, the~matrix-valued Faddeev amplitude for a baryon of total momentum $P= p_q + p_d$, which~expresses the relative momentum correlation between the dressed-quarks and -diquarks within the baryon.  The~shaded rectangle demarcates the kernel of the Faddeev equation: \emph{single line}, dressed-quark propagator; $\Gamma$,  diquark correlation amplitude; and \emph{double line}, diquark propagator.}
\label{figFaddeev}
\end{figure}

The existence of these strong correlations between two dressed quarks is the path to converting the three valence-quark bound-state problem into the simpler Faddeev equation drawn in Figure\,\ref{figFaddeev}, without~loss of dynamical information~\cite{Eichmann:2009qa}.
The three-gluon vertex, a~definitive feature of QCD's non-Abelian nature, is not an explicit part of the scattering kernel in this picture.  Instead, one profits from the following features,
phase-space factors materially enhance two-body interactions over $n\geq 3$-body interactions and
the primary three-body force, produced by a three-gluon vertex attaching once, and~only once, to~each valence quark, vanishes when projected into the colour-singlet channel;
and subsequently capitalises on diquark dominance in the two-body subsystems.
Then, while~iterated, overlapping three-body terms might alter fine details of baryon structure, the~primary effect of non-Abelian multi-gluon vertices is manifested in the formation of diquark correlations. (On the other hand, the~three-gluon vertex appears to play a material and distinguishable role in the formation of hybrid hadrons~\cite{Xu:2018cor} and glueballs~\cite{Souza:2019ylx}).  Accordingly, the~active kernel describes binding in the baryon through diquark breakup and reconstitution, mediated by exchange of a dressed-quark.  The~properties and interactions of such a baryon are chiefly determined by the quark$+$diquark structure evident in Figure\,\ref{figFaddeev}.

The spectrum of baryons produced by the Faddeev equation~\cite{Eichmann:2016hgl, Lu:2017cln, Chen:2019fzn, Yin:2019bxe} is like that found in the three-constituent quark model and consistent with lQCD analyses~\cite{Edwards:2011jj}.  Notably, modern data and recent analyses have reduced the number of missing resonances~\cite{Ripani:2002ss, Burkert:2012ee, Kamano:2013iva, Crede:2013sze, Mokeev:2015moa, Anisovich:2017pmi}.

\subsection{Proton Wave~Function}
\begin{figure}[H]
\hspace*{-2.5em}\begin{tabular}{ccc}
\parbox[c]{0.31\linewidth}{\includegraphics[clip,width=\linewidth]{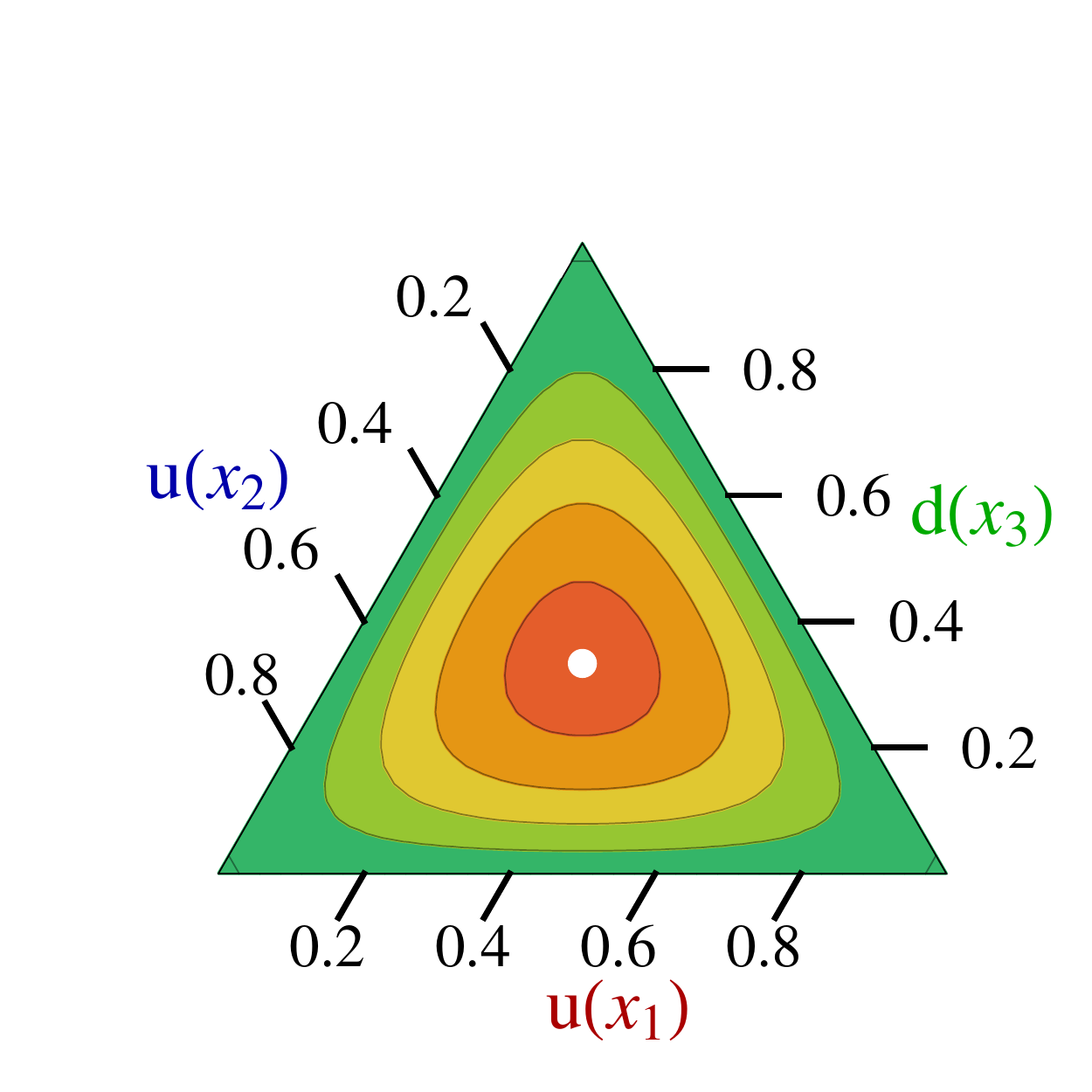}} &
\parbox[c]{0.31\linewidth}{\includegraphics[clip,width=\linewidth]{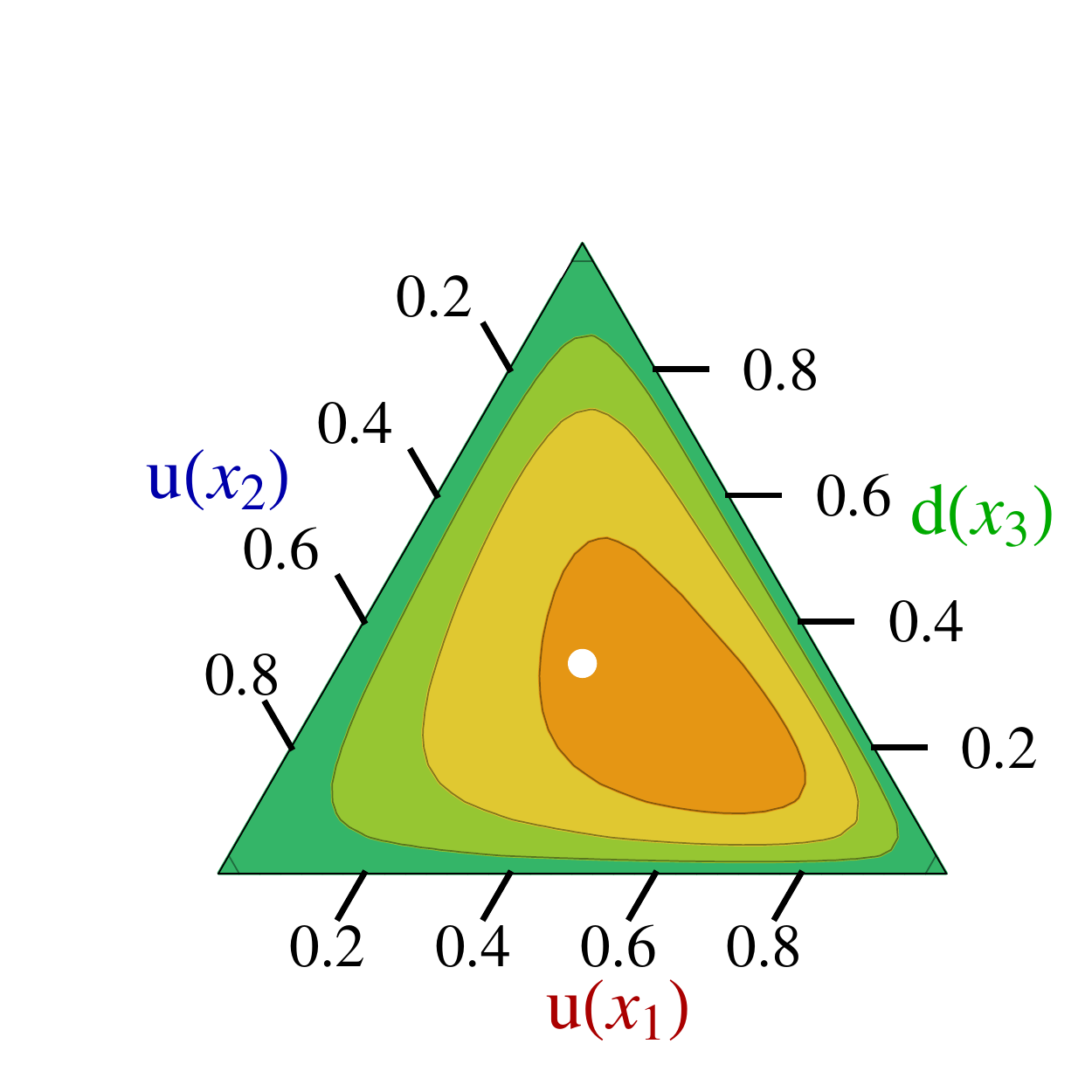}} &
\parbox[c]{0.37\linewidth}{\includegraphics[clip,width=\linewidth]{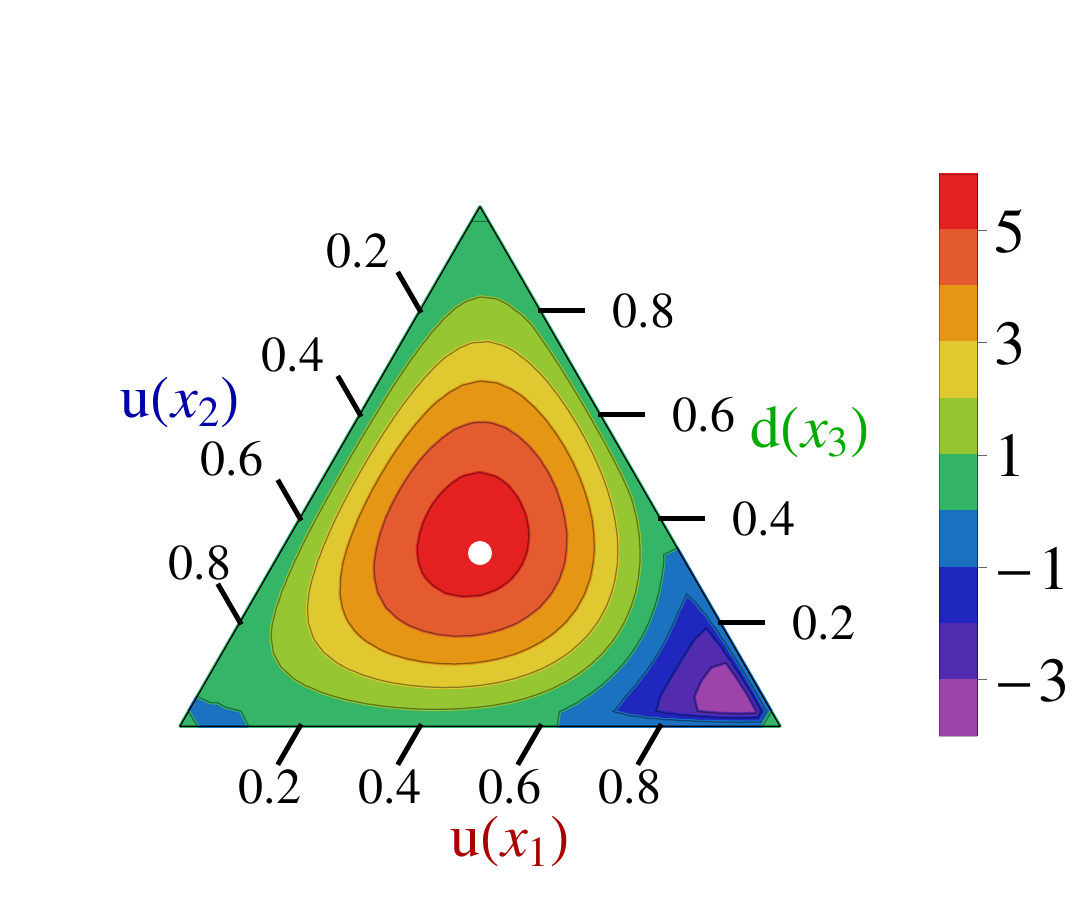}}
\end{tabular}\vspace*{-2ex}
\caption{
Baryon leading-twist dressed-quark distribution amplitudes depicted using barycentric plots.
\emph{Left panel}---asymptotic profile, baryon PDA, $\varphi_N^{\rm asy}([x])=120 x_1 x_2 x_3$; \emph{middle panel}---computed~proton PDA evolved to $\zeta=2\,$GeV, which~peaks at \mbox{$([x])=(0.55,0.23,0.22)$}; and \emph{right panel}---Roper resonance PDA.  The~white circle in each panel marks the centre of mass for $\varphi_N^{\rm asy}([x])$, whose peak lies at \mbox{$([x])=(1/3,1/3,1/3)$}.
Here, $x_1, x_2, x_3$ indicate the fraction of the bound-state's light-front momentum carried by the associated quark; naturally, $x_1+x_2+x_3=1$.  The~amplitudes are dimensionless; hence, the~height is simply a real number.
}
\label{PlotPDAs}
\end{figure}

After solving the Faddeev equation in Figure\,\ref{figFaddeev} for the proton's Faddeev amplitude, one can then compute the proton's dressed-quark leading-twist PDA~\cite{Mezrag:2017znp}.  The~result for this proton ``wave function'' is depicted in Figure\,\ref{PlotPDAs}.

Table\,\ref{interpolation} lists the proton PDA's four lowest-order moments.  They reveal important insights, e.g.,\ when the proton is pictured as solely a quark+sca\-lar-diquark correlation, $\langle x_2 \rangle_u=\langle x_3 \rangle_d$, because~these two form the scalar quark+quark correlation, and the system is very asymmetric, with~the PDA's peak being shifted strongly in favour of $\langle x_1 \rangle_u > \langle x_2 \rangle_u$.  This~outcome conflicts with lQCD \mbox{results~\cite{Braun:2014wpaS, Bali:2015ykx}}.  On~the other hand, as~explained above, realistic Faddeev equation calculations suggest that pseudovector diquark correlations are an essential piece of the proton's wave function.  When these $\{uu\}$ and $\{ud\}$ diquarks are included, momentum is shared more evenly, shifting from the spectator $u(x_1)$ quark into $u(x_2)$, $d(x_3)$.  Adding these diquarks with the known weighting, the~PDA's peak shifts back toward the centre, relocating to $([x])=(0.55,0.23,0.22)$, and the computed values of the first moments line up with those computed using lQCD.  This~convergence delivers a more complete understanding of the lQCD simulations, which~by that means are seen to confirm a picture of the proton as a bound-state with both tight scalar \emph{and} pseudovector diquark correlations, wherein~the scalar diquarks are responsible for $\approx 65$\% of the Faddeev amplitude's canonical~normalisation.

Importantly, like ground-state $S$-wave mesons~\cite{Chang:2013pqS, Shi:2014uwaS, Braun:2015axa, Gao:2016jka, Zhang:2017bzy, Chen:2017gck}, the~leading-twist PDA of the ground-state nucleon is both broa\-der than $\varphi_N^{\rm asy}([x])$ (defined by Table\,\ref{interpolation}, Row~1) and decreases monotonically away from its maximum in all directions, i.e.,\ the~PDAs of these ground-state $S$-wave systems possess endpoint enhancements, but~neither humps nor bumps.
Models which produce such humped structures were previously considered reasonable~\cite{Chernyak:1983ej}.  However, it is now clear that pointwise behaviour of this type is in conflict with QCD.  The~models may nonetheless be viewed as possessing a qualitative truth, insofar as they represent a way by which endpoint enhancements can be expressed in hadron PDAs if one limits oneself to the polynomial basis characterising QCD in the neighbourhood $\Lambda_{\rm QCD}/\zeta \simeq 0$.

As with the pion prediction described in Section\,\ref{secPionFF}, the~veracity of this result for the proton PDA can be tested in future experiments.  For~instance, it can be used to provide the first realistic evaluation of the scale at which exclusive measurements involving proton targets may reasonably be compared with predictions based on pQCD hard scattering formulae.  Analogous to the pion case, the~value of this mass-scale is an empirical manifestation of the emergence of mass, here within the three-valence-quark proton~bound-state.

\begin{table}[H]
\caption{
%
Computed values of the first four moments of the proton and Roper-resonance PDAs.  The~error on $f_N$, a~dynamically-determined quantity which measures the proton's ``wave function at the origin'', reflects a nucleon scalar diquark content of $65\pm 5$\%, and values in rows marked with ``$\not\supset \mbox{av}$'' were obtained assuming the baryon is constituted solely from a scalar diquark.
Including axial-vector diquark correlations, the~mean absolute relative difference between continuum and lattice results improves by 36\%.
(All results listed at a renormalisation scale $\zeta=2\,$GeV.)
}
\label{interpolation}
\centering
\begin{tabular}[t]{lcccc}
\toprule
 & \boldmath $10^3 f_N/\mbox{GeV}^2$ & \boldmath $\langle x_1\rangle_u$ & \boldmath $\langle x_2\rangle_u$ & \boldmath $\langle x_3\rangle_d$ \\\midrule
{asymptotic PDA}
 & & $0.333\phantom{(99)}$ & $0.333\phantom{(9)}$ & $0.333\phantom{(9)}$ \\\midrule
lQCD \mbox{\cite{Braun:2014wpaS}} & {$2.84~(33)$} & $0.372~(7)\phantom{9}$ & 0.314~(3) & 0.314~(7) \\
lQCD \mbox{\cite{Bali:2015ykx}} & $3.60~(6)\phantom{9}$ &  $0.358~(6)\phantom{9}$ & $0.319~(4)$ & 0.323~(6) \\\midrule
DSE proton \mbox{\cite{Mezrag:2017znp}}& $3.78~(14)$ & $0.379~(4)\phantom{9}$ & 0.302~(1) & 0.319~(3) \\
DSE proton $\not\supset \mbox{av}$  & $2.97\phantom{(17)}$ & $0.412\phantom{(17)}$ & $0.295\phantom{(7)}$ & $0.293\phantom{(7)}$ \\\midrule
DSE Roper \mbox{\cite{Mezrag:2017znp}}   & $5.17~(32)$ & $0.245~(13)$ & $0.363~(6)$ & $0.392~(6)$ \\
DSE Roper $\not\supset \mbox{av}$   & $2.63\phantom{(14)}$ & $0.010\phantom{(19)}$ & $0.490\phantom{(9)}$ & $0.500\phantom{(9)}$ \\
\bottomrule
\end{tabular}
\end{table}
\unskip

\subsection{Proton's First Radial~Excitation}
Almost immediately after discovery of the Roper resonance~\cite{Roper:1964zza, BAREYRE1964137, AUVIL196476, PhysRevLett.13.555, PhysRev.138.B190}, the~proton's first positive-parity excitation, questions were asked regarding the nature of like-parity excitations of ground-state positive-parity baryons.~A~lucid picture is now emerging following~\cite{Burkert:2017djo},
(\emph{i})~the~acquisition and analysis of a large amount of very precise nucleon-resonance electroproduction data, involving~single- and double-pion final states on a large domain of energy and momentum-transfer;
(\emph{ii})~development of an advanced dynamical reaction theory, able to simultaneously describe all partial waves extracted from available data;
(\emph{iii}) formulation and extensive application of the DSE approach to the continuum bound state problem in relativistic quantum field theory, which~expresses diverse local and global impacts of EHM; and
and \emph{(iv}) the improvement of constituent quark models so that they can also qualitatively incorporate these features of strong QCD.
In this picture, such states are primarily radial excitations of the associated ground-state baryon, comprised of a well-defined dressed-quark core supplemented by a meson~cloud.

Using the Faddeev equation framework sketched above, properties of the Roper-resonance's dressed-quark core have been exposed~\cite{Chen:2017pse, Segovia:2015hraS}: it is found that the scalar functions in the Roper's Faddeev amplitude have a zero at quark+diquark relative momentum $\surd \ell^2 \approx 0.4\,$GeV$\approx 1/[0.5\,{\rm fm}]$.  Working with this input, the authors of \,\cite{Mezrag:2017znp} delivered the associated leading-twist PDA, depicted in the rightmost panel of Figure\,\ref{PlotPDAs} and whose first four moments are listed in Table\,\ref{interpolation}.  The~prediction reveals some interesting features, e.g.,
the PDA of this excitation is not positive definite and there is a conspicuous locus of zeros in the lower-right section of the barycentric plot, both of which duplicate aspects of the wave function describing the first radial excitation of a quantum mechanical system that are also seen in the PDAs of meson radial excitations~\cite{Li:2016dzv, Li:2016mah},
and here the influence of pseudovector diquarks is contrary to that in the ground-state, viz.\ they shift momentum into $u(x_1)$ from $u(x_2)$, $d(x_3)$.

These observations highlight that the ground state is just one isolated member of a set of Hamiltonian eigenvectors with infinitely many elements:~many Hamiltonians can possess practically equivalent ground states and yet lead to excited-state spectra that are vastly different.  Moreover, masses~alone, as~infrared-dominated quantities, contain relatively little information.  Different~Hamiltonians may adequately reproduce known hadron spectra; but these same Hamiltonians can yield predictions that disagree markedly when used to compute structural properties.  Such properties---like wave functions and the $Q^2$-dependence of elastic and transition form factors---possess the greatest discriminating power.   Therefore, study of the structure of nucleon resonances is a critical complement to that of ground-state nucleons and mesons because it is capable of revealing additional novel features of strong QCD.  Modern theory must be deployed to compute observable properties of all these systems.  Aspects of this effort are sketched elsewhere~\cite{Brodsky:2020vco, Aznauryan:2012baS}.

\subsection{Emergent Features of Nucleon Form~Factors}
The character of emergent mass and the diquark correlations within baryons it induces are visible in baryon elastic and transition form factors, and particular examples of contemporary significance are neutron and proton elastic form factors.  Nucleons are the most basic elements of nuclear physics; and these form factors are manifestations of the nature of the nucleons' constituents and the dynamics holding them~together.

Paradigmatic shifts in our understanding of these things are being driven by new, precise form factor data. This is nowhere more evident than in analyses of experimental data acquired in the past vicennium, which~have established a new ideal.  Namely, despite its elementary valence-quark content, the~nucleon's internal structure is very complex.  For~instance, there are measurable differences between the distributions of charge and magnetisation throughout the interiors~\cite{Jones:1999rz} and between the way these qualities are carried by the different quark flavours~\cite{Cates:2011pz, Segovia:2014aza, Segovia:2016zyc}.  The~aim now is to explain the observations in terms of emergent features of the strong~interaction.

In this connection, it is here worth highlighting that the behaviour of the proton's electric form factor on $Q^2 \gtrsim 4\,$GeV$^2$ is particularly sensitive to the running of the dressed-quark mass (displayed in Figure\,\ref{gluoncloud}), especially the rate at which the dressed-quark mass runs between the infrared and ultraviolet domains~\cite{Wilson:2011aa, Segovia:2014aza, Cui:2020rmuS}.

The proton's momentum-space charge and magnetisation distributions are expressed in combinations of the two Poincar\'e-invariant elastic form factors that are needed to express the proton's electromagnetic current:
\begin{equation}
i e \, \bar u(p^\prime) \big[ \gamma_\mu F_1(Q^2) +
\frac{Q_\nu}{2m_N}\, \sigma_{\mu\nu}\,F_2(Q^2)\big] u(p)\,,
\end{equation}
where $m_N$ is the nucleon mass; $Q=p^\prime - p$, $u(p)$ and $\bar u(p^\prime)$ are, respectively, spinors describing the incident and scattered proton; and~$F_{1,2}(Q^2)$ are the proton's Dirac and Pauli form factors.  The~combinations that feature in the electron-proton elastic scattering cross  section are the charge and magnetisation distributions~\cite{Sachs:1962zzc}
\begin{equation}
\label{GEMpeq}
G_E(Q^2)  =  F_1(Q^2) - (Q^2/[4 m_N^2]) F_2(Q^2)\,,\quad
G_M(Q^2) =  F_1(Q^2) + F_2(Q^2)\,.
\end{equation}

Data available before the year 1999 led to a view that
\begin{equation}
\label{GEeqGM}
\left. \mu_p\, \frac{G_E^p(Q^2)}{G_M^p(Q^2)} \right|_{\rm Rosenbluth} \approx 1\,,
\end{equation}
$\mu_p\,G_M^p(Q^2=0)=1$; therefore, a~conclusion that the distributions of charge and magnetisation within the proton are approximately identical~\cite{Perdrisat:2006hj}.  At~the time, the~proton was viewed as a simple bound state, wherein quark+quark correlations and attendant orbital angular momentum played little role.  Equation\,\eqref{GEeqGM} is consistent with this~picture.

The situation changed dramatically when the combination of high-energy, -current and -polarisation at JLab enabled a new type of experiment to be performed, viz.\ polarisation-transfer reactions~\cite{Jones:1999rz}, which~are directly proportional to $G_E(Q^2)/G_M(Q^2)$ \cite{Akhiezer:1974em, Arnold:1980zj}.~A~series of these experiments~\cite{Jones:1999rz, Gayou:2001qd, Punjabi:2005wq, Puckett:2011xgS, Puckett:2010ac} has determined that
\begin{equation}
\label{GEGMJLab}
\left. \mu_p\, \frac{G_E^p(Q^2)}{G_M^p(Q^2)} \right|_{\rm JLab\,PT}  \approx 1 - {\rm constant} \times Q^2\,,
\end{equation}
where the constant is such that the ratio might become negative for $Q^2 \gtrsim 8\,$GeV$^2$.  This~behaviour contrasts starkly with Equation\,\eqref{GEeqGM}; and since the proton's magnetic form factor is reliably known on a spacelike domain that extends to $Q^2 \approx 30\,$GeV$^2$ \cite{Kelly:2004hm, Bradford:2006yz}, the~$Q^2$-dependence of $G_E^p/G_M^p$ exposes novel features of the proton's charge distribution, as~expressed in $G_E^p(Q^2)$.

An understanding of the behaviour in Equation\,\eqref{GEGMJLab} is provided by the analyses in \mbox{ \,\cite{Wilson:2011aa, Segovia:2014aza, Cui:2020rmuS}}, and the answer lies largely with the proton's Pauli form factor.  $F_2^p$ is a gauge of the proton's magnetisation distribution. Ultimately, this magnetisation is borne by the dressed-quarks and influenced by the $k^2$-dependence of (\emph{i}) their propagators and (\emph{ii}) the correlations amongst them.  Both are expressed in the Faddeev~wavefunction.

Thus, for~the sake of argument,~suppose that dressed-quarks are described by a momentum- \linebreak independent mass-function, e.g.,\ as obtained using a symmetry-preserving regularisation of a vector$\,\times\,$vector contact interaction~\cite{GutierrezGuerrero:2010md}.  They then behave as Dirac particles with constant Dirac values for their magnetic moments.  Consequently, the~composite proton possesses a hard Pauli form factor and this produces a zero in $G_E^p$ at $Q^2 \approx 4\,$GeV$^2$ \cite{Wilson:2011aa, Frank:1995pv}.

Alternatively, suppose that the dressed-quarks have a momentum-dependent mass-function, like~that depicted in Figure\,\ref{gluoncloud}, which~is large at small momenta but vanishes as their momentum increases.  At~infrared momenta the dressed-quarks will then act as constituent-like particles with a large magnetic moment, but as the probe momentum grows, their mass and magnetic moment will drop toward zero.  (\emph{N.B}.\ Massless fermions do not have a measurable magnetic moment~\cite{Chang:2010hb}, so that any significant magnetic moment for a constituent-like quark is an emergent feature.)  Such~dressed-quarks will produce a proton Pauli form factor that is large on $Q^2 \simeq 0$ but drops rapidly on the domain of transition between the infrared and ultraviolet domains, to~give a very small result at large-$Q^2$.  The~proton's Dirac form factor is far less sensitive to spin-related effects; hence the interplay between the Dirac and Pauli form factors, expressed in Equation\,\eqref{GEMpeq}, entails that $G_E(Q^2)$ must have a zero at larger values of $Q^2$ when calculated with a running mass function than we computed with momentum-independent dressed-quark masses.
The precise form of the $Q^2$-dependence will depend on the evolving nature of the angular momentum correlations between the dressed-quarks~\cite{Segovia:2014aza, Cui:2020rmuS}.

The Class-C DSE analyses in \,\cite{Segovia:2014aza, Cui:2020rmuS} implement a dressed-quark mass function that is distinguished by a particular transition rate between the non-perturbative and perturbative domains.  If~that rate were increased, then the transformation to partonic quarks would become more rapid; hence~the proton's Pauli form factor would drop even more quickly to zero.  In~this event, the~quark angular momentum correlations, embodied in the diquark structure, remain but the separate dressed-quark magnetic moments diminish markedly.  Thus a quicker transition will push the zero in $\mu_p G_{E}^p/G_{M}^p$ to larger values of $Q^2$.  Moreover, there will be a rate of transformation beyond which the zero disappears completely: there is no zero in a theory whose mass function rapidly becomes partonic.  (For instance, pQCD analyses cannot produce a zero.)  These expectations are realised in explicit calculations, as~illustrated in Figure~4 of Ref.\,\cite{Segovia:2014aza}.

It follows that the possible existence and location of a zero in the ratio of proton elastic form factors $\mu_p G_{E}^p(Q^2)/G_{M}^p(Q^2)$ are a fairly direct measure of the character of EHM in the SM.  Consequently, in~pushing experimental measurements of this ratio, and~thereby the proton's electric form factor, to~larger momenta, i.e.,\ in reliably determining the proton's charge distribution, there is an exceptional opportunity for a positive dialogue between experiment and theory.  That feedback should assist greatly with contemporary efforts to reveal the character of the SM's strong interaction and its emergent~phenomena.

\begin{figure}[t]
\begin{center}
\begin{tabular}{lr}
\includegraphics[clip,width=0.425\linewidth]{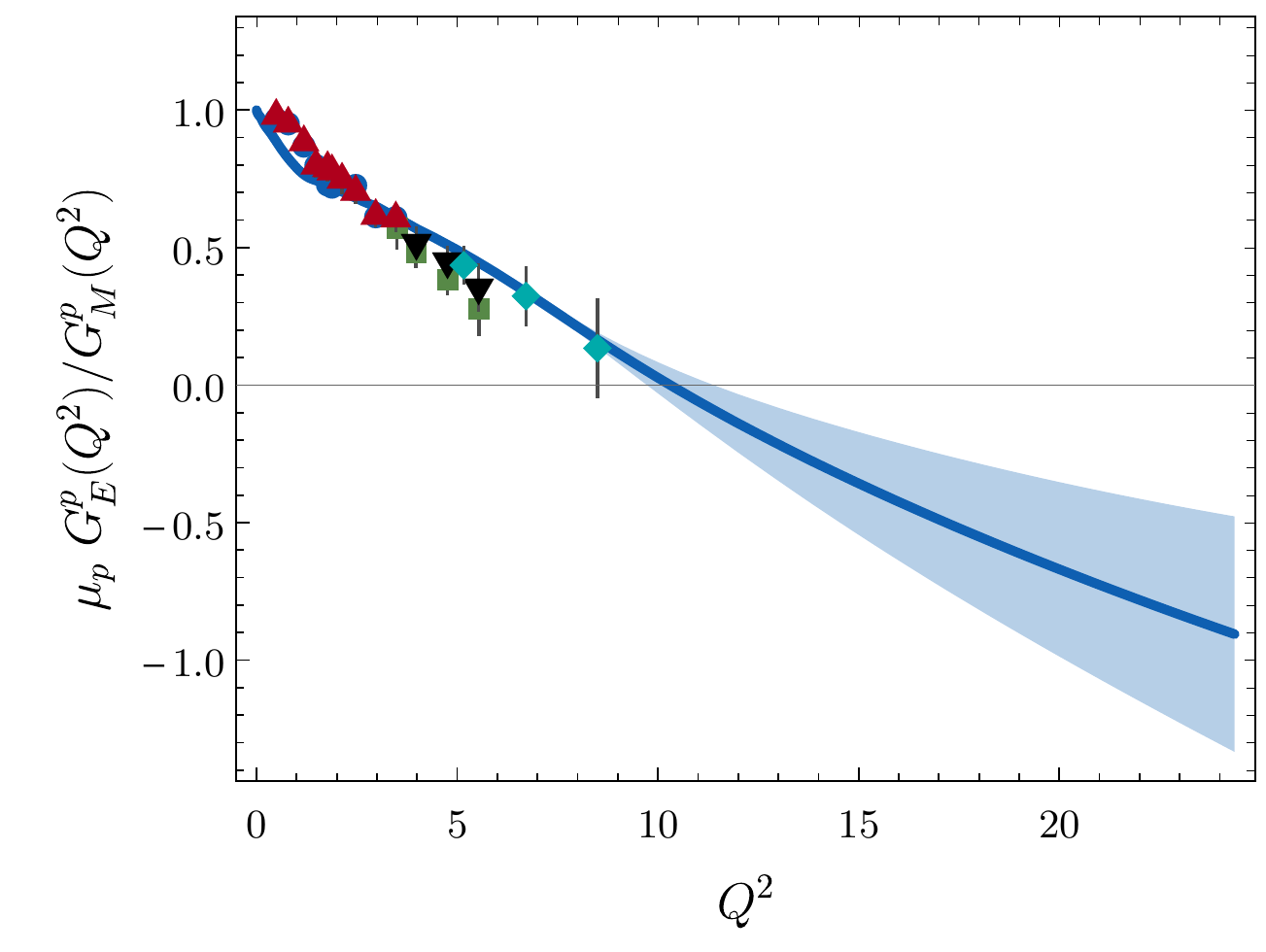}\hspace*{2ex } &
\includegraphics[clip,width=0.425\linewidth]{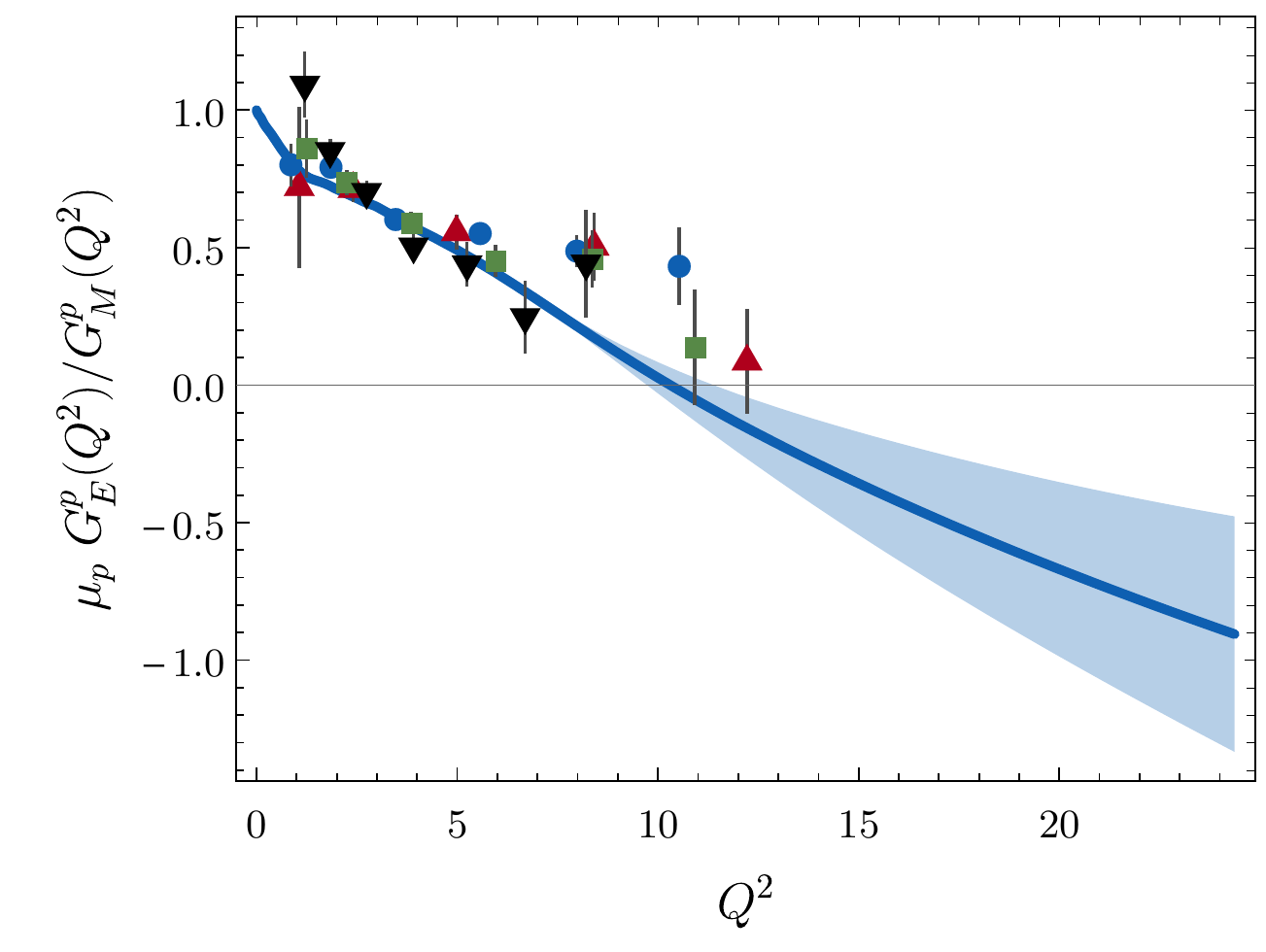}\vspace*{-0ex}
\end{tabular}
\begin{tabular}{lr}
\includegraphics[clip,width=0.425\linewidth]{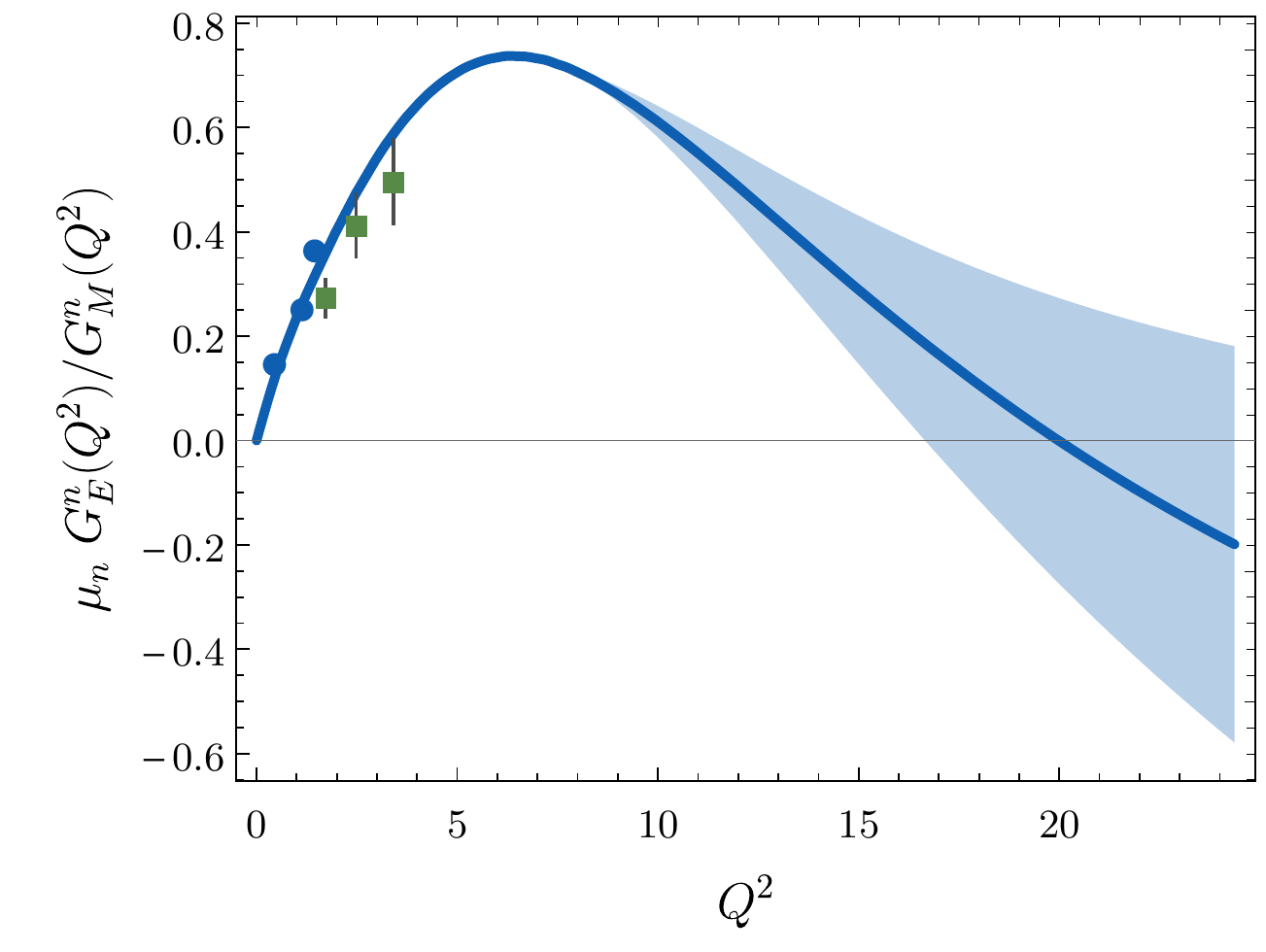}\hspace*{2ex } &
\includegraphics[clip,width=0.425\linewidth]{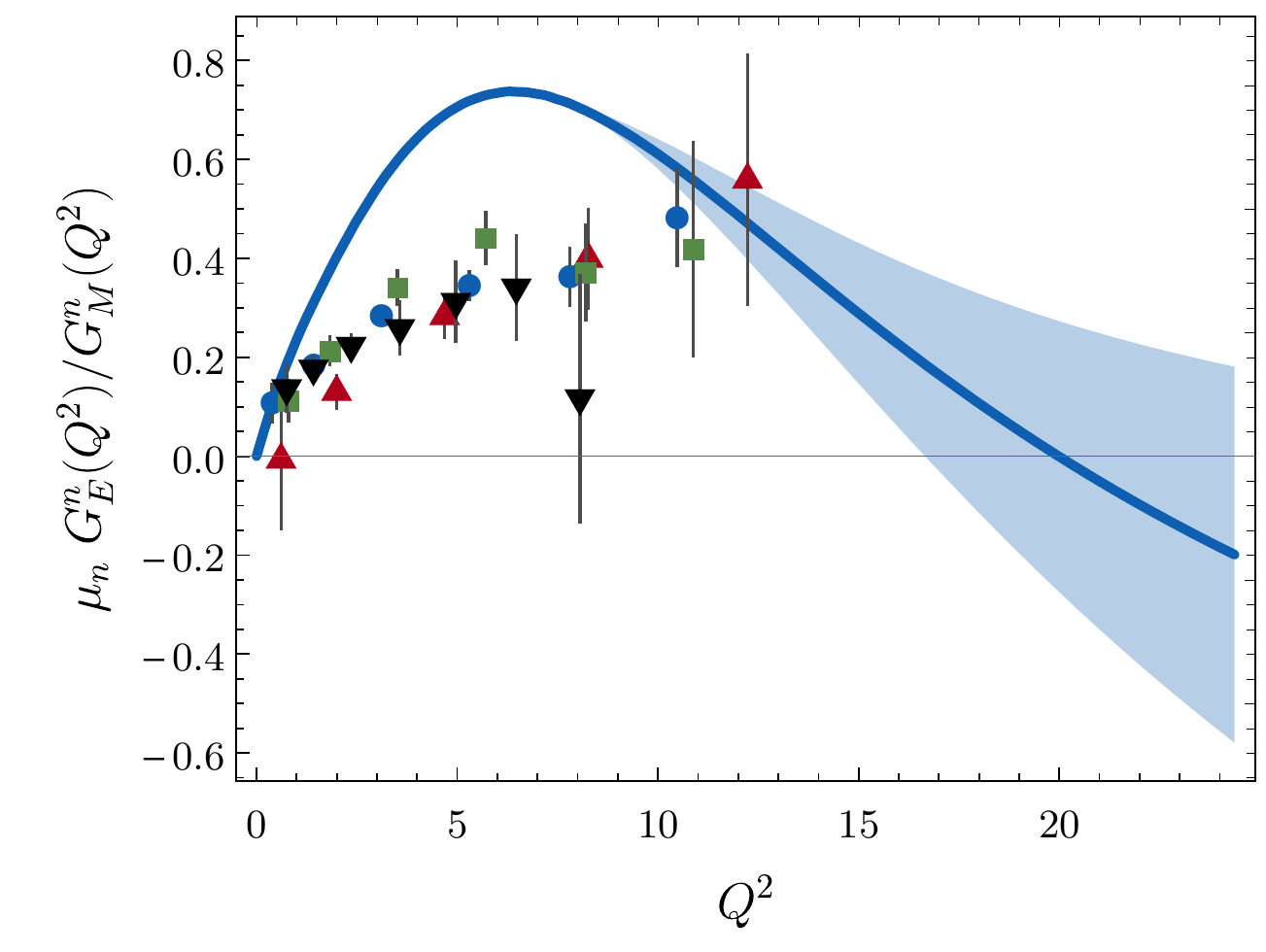}\vspace*{-1ex}
\end{tabular}
\end{center}
%
\caption{
{Ratios of Sach's form factors}, $\mu_N G_E^N(Q^2)/G_M^N(Q^2)$.
\emph{Upper panels}--- Proton.  \underline{{Left}} 
, \,\cite{Cui:2020rmuS} calculation compared with data
(red up-triangles~\cite{Jones:1999rz};
green squares~\cite{Gayou:2001qd};
blue circles~\cite{Punjabi:2005wq};
black down-triangles~\cite{Puckett:2011xgS};
and cyan diamonds~\cite{Puckett:2010ac});
\underline{{right}}, compared with available lQCD results, drawn from Ref.\,\cite{Kallidonis:2018cas}.
\emph{Lower panels}---Neutron.  \underline{{Left}}, comparison with data (blue circles~\cite{Madey:2003av} and green squares~\cite{Riordan:2010idS}); \underline{{right}}, with~available lQCD results, drawn from the work in \,\cite{Kallidonis:2018cas}.
In all panels, the~$1\sigma$ band for the SPM approximants is shaded in light blue.
}
\label{GEonGM}
\end{figure}

Given such potential, the~analysis in \,\cite{Segovia:2014aza} was recently revisited~\cite{Cui:2020rmuS}, using improved algorithms to calculate the form factors on $Q^2 \gtrsim 10\,m_p^2$.  The~new study employed a statistical implementation of the Schlessinger point method (SPM)  \cite{Binosi:2019ecz,Chen:2018nsg, Schlessinger:1966zz, PhysRev.167.1411, Tripolt:2016cya, Binosi:2018rht} for the interpolation and extrapolation of smooth functions to deliver predictions for form factors on $Q^2>9\,m_p^2$ with a quantified uncertainty estimate.  Results for the form factor ratios discussed herein are presented in Figure\,\ref{GEonGM}.

For the proton, the~new analysis predicts
\begin{equation}
\mu_p\, \frac{G_E^p(Q_{z_p}^2)}{G_M^p(Q_{z_p}^2)} = 0\,, \quad Q_{z_p}^2 = 10.3^{+1.1}_{-0.7}\,{\rm GeV}^2\,.
\end{equation}
This value is compatible with, although~a little larger than, that obtained earlier~\cite{Segovia:2014aza}: $Q_{z_p}^2 \approx 9.5\,$GeV$^2$; and a more recent inference based on $\rho$-meson elastic form factors~\cite{Xu:2019ilh}: $Q_{z_p}^2 \approx 9.4(3)\,$GeV$^2$.

Regarding the neutron, the authors of \,\cite{Segovia:2014aza} predicted a peak in this ratio at $Q^2\approx 6\,$GeV$^2$, which~is reproduced in \,\cite{Cui:2020rmuS}.  Furthermore, it located a zero at $Q_{z_n}^2\approx 12\,$GeV$^2$.  With~the statistical SPM method for reaching to large-$Q^2$:
\begin{equation}
 \mu_n\, \frac{G_E^n(Q_{z_n}^2)}{G_M^n(Q_{z_n}^2)} = 0\,, \quad Q_{z_n}^2 = 20.1^{+10.6}_{-\phantom{1}3.5}\,{\rm GeV}^2\,,
\end{equation}
viz.\ at $1\sigma$ SPM confidence level, this ratio is likely to exhibit a zero, but~it probably lies beyond the reach of 12\,GeV beams at JLab.  On~the other hand, the~prediction of a peak in $R_{EM}^n=\mu_n G_E^n/G_M^n$, which~is a harbinger of the zero in this ratio, can be tested at the 12\,GeV~JLab.

The properties of the dressed-quark propagators and bound-state amplitudes which influence the appearance of a zero in $R_{EM}^n$ are qualitatively the same as those described in connection with $R_{EM}^p$.  However, because~of the different electric charge weightings attached to the quark contributions in the neutron (1 valence $u$-quark and 2 valence $d$-quarks), the~quantitative effect is opposite to that for the proton (2 valence $u$-quarks and 1 valence $d$-quark).  Namely, when the transformation from dressed-quark to parton is accelerated, the~zero occurs at smaller $Q^2$.  On~the other hand, a~model which generates a momentum-independent dressed-quark mass typically produces no zero in the neutron ratio~\cite{Wilson:2011aa}.

In order to understand the source of these features, consider that the strange quark contributes very little to nucleon electromagnetic form factors~\cite{Aniol:2005zf, Armstrong:2005hs}, in~which case one can write
\begin{equation}
\label{eqFlavourSep}
G_{E}^p = e_u G_E^{p,u} - |e_d| G_E^{p,d} \,,\;
G_{E}^n = e_u G_E^{n,u} - |e_d| G_E^{n,d}
\end{equation}
($e_u=2/3$, $e_d = -1/3$),
where the contribution from each quark flavour is made explicit.  Consider next that charge-symmetry is almost exact in QCD; to wit,
\begin{equation}
G_E^{n,d} = G_E^{p,u}\,, \quad G_E^{n,u} = G_E^{p,d}\,.
\end{equation}
Therefore, to~a very good level of approximation,
\begin{equation}
G_{E}^n = e_u G_E^{n,u} - |e_d| G_E^{n,d} =  e_u G_E^{p,d} - |e_d| G_E^{p,u}\,.
\label{GEnCS}
\end{equation}

Now, with~a zero in $G_{E}^p$ at $Q^2 = Q_{z_p}^2=:s_z^p$, one has $G_E^{p,d}(s_z) = 2 \,G_E^{p,u}(s_z)$ and hence \mbox{$G_{E}^n(s_z) = G_E^{p,u}(s_z)>0$.}  This~shows that although the behaviours of $G_E^{p,u}$ and  $G_{E}^p$ are qualitatively similar, the~zero in $G_E^{p,u}$ occurs at a larger value of $Q^2$ than that in $G_{E}^p$ itself.  Under~these conditions, any~zero in $G_{E}^n$ must occur at a larger value of $Q^2$ than the zero in $G_E^p$, a~prediction confirmed in~Figure\,\ref{GEonGM}.

\begin{figure}[H]
\centering
\includegraphics[clip,width=0.60\linewidth]{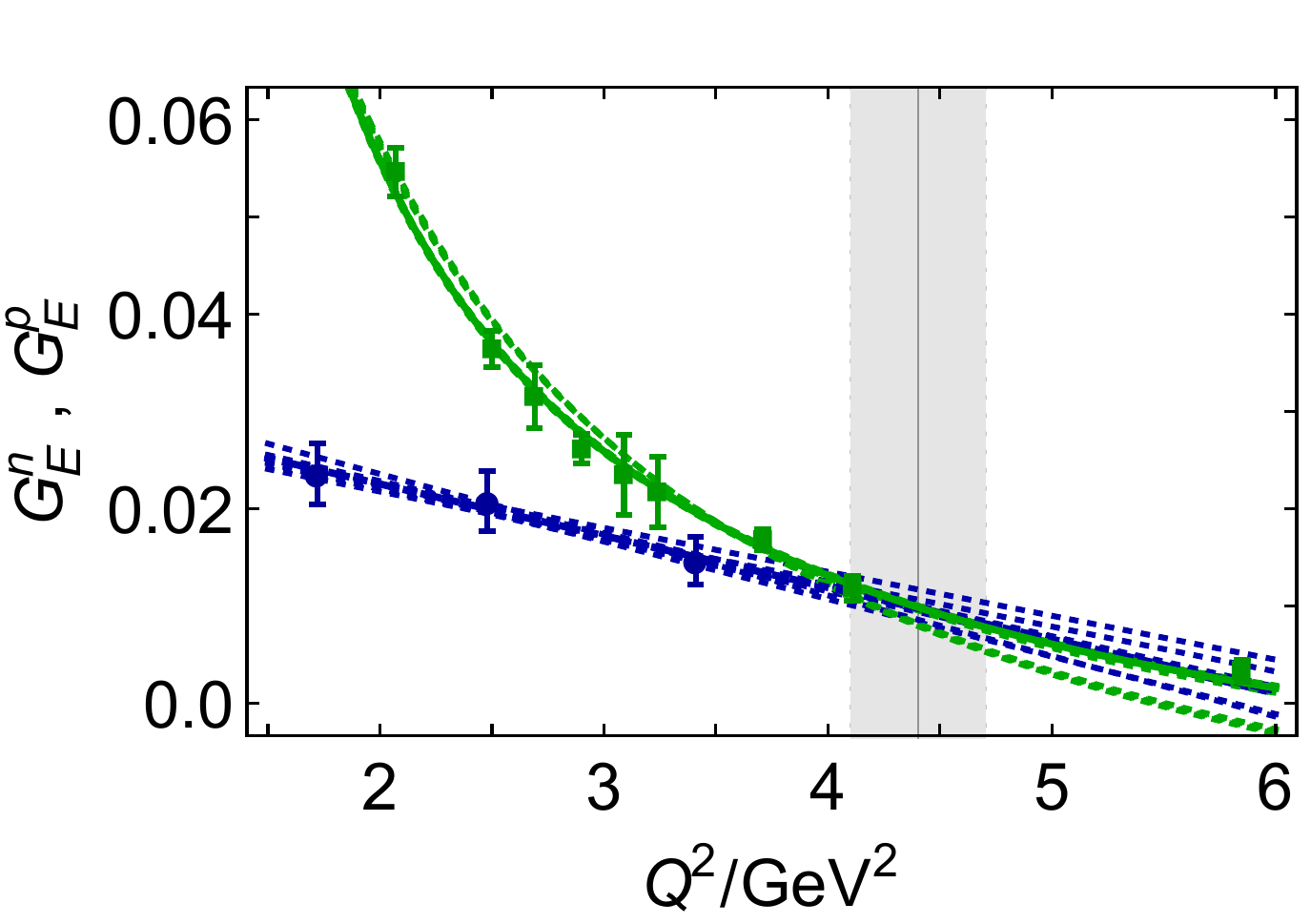}
%
\caption{
Data:
neutron electric form factor---blue circles~\cite{Riordan:2010idS},
and
proton electric form factor---green squares~\cite{Arrington:2007ux}.
Associated curves depict least-squares fits obtained via a jackknife analysis of the depicted $G_E^n$, $G_E^p$ data.
Vertical black line within grey bands marks the boundary of the domain described by Equation\,\eqref{GEnbigger}.
}
\label{GEnequalsGEp}
\end{figure}

A curious effect follows from the combination of faster-than-dipole decrease of the proton's electric form factor (and probable appearance of a zero) and steady increase of $R_{EM}^n$ to a peak at $Q^2\approx 6\,$GeV$^2$.  Namely, there must be a domain of larger $Q^2$ upon which the neutron's electric form factor is greater than that of the proton's; therefore, at~some value of $Q^2$, the~electric form factor of the neutral compound fermion is actually larger than that of its positive counterpart.
This was predicted in \,\cite{Segovia:2014aza} and confirmed by the updated analyses in \,\cite{Cui:2020rmuS}.  It is therefore worth testing with available data~\cite{Arrington:2007ux, Riordan:2010idS}.  Figure~\ref{GEnequalsGEp} depicts $G_E^n$, $G_E^p$ data along with a set of associated least-squares fits produced using a jackknife analysis.  On~the domain currently covered by data, $G_E^n < G_E^p$; but~extrapolation of the fits suggests that
\begin{equation}
\label{GEnbigger}
G_E^n(Q^2) \stackrel{Q^2 > Q_{n>p}^2}{>} G_E^p(Q^2)\,,\quad Q_{n>p}^2 = 4.4(3)\,{\rm GeV}^2.
\end{equation}
With 12\,GeV operations under way at JLab, neutron electric form factor data to $Q^2=10.2\,$GeV$^2$ will soon be available~\cite{E12-09-016}, enabling this prediction to be~tested.

It is worth reiterating that the existence and location of a zero in $G_E^p$---similarly for $G_E^n$--- \linebreak are~sensitive to the rate at which the dressed-quark mass function transits from the strong to perturbative QCD domains: denote this rate by $R_M^{sp}$.  Note, too, that in contrast to $G_E^{p,u}$, $G_E^{p,d}$ evolves more slowly with changes in $R_M^{sp}$.  This~inertia derives from the $d$-quark being preferentially contained within a (soft) scalar diquark.  Subject to these insights, consider Equation\,\eqref{GEnCS}: with the location of a zero in $G_E^{p,d}$ shifting slowly to larger values of $Q^2$, but that in $G_E^{p,u}$ moving rapidly (as noted above, the~zero in $G_E^{p,u}$ disappears if $R_M^{sp}$ is sufficiently quick) one is subtracting from $G_E^{p,d}(Q^2)$ a function whose domain of positive support is increasing in size.  That operation typically shifts the zero in $G_E^n$ to smaller values of $Q^2$, potentially enabling a zero in $G_E^n$ even when that in $G_E^p$ has~disappeared.

\section{Epilogue}
The Lagrangian that defines QCD, the~strong interaction sector of the Standard Model (SM), appears very simple, yet it is responsible for a remarkable array of high-level phenomena with enormous apparent complexity.  One can argue that the foundation for all such effects is laid by the emergence of hadronic mass (EHM), whose consequences are variously expressed in a wide range of empirical observables.  This~emergence itself is currently beyond a reductionist explanation, unless~one is content to find that in the need for ultraviolet renormalisation of four-dimensional quantum field theory.  Even supposing the latter, then the size of the associated mass-scale, $\Lambda_{\rm QCD}$, is not something which can be predicted from within the~SM.

The simplest expression of EHM is to be found in the generation of a running mass for the gluon.  Mechanically, this is driven by gluon self-interactions, and the scale of the effect is known once $\Lambda_{\rm QCD}$ is fixed.  It is interesting theoretically to explore the impact on observables induced by reducing/increasing $\Lambda_{\rm QCD}$.  Perhaps there is a critical value (or set of values) at which the Universe would appear very different?

Gluon mass generation in the SM entails that quarks, which~are massless in the absence of a Higgs mechanism, also acquire a running mass, whose scale at infrared momenta is roughly one-third of the proton mass.  This~effect is known as dynamical chiral symmetry breaking (DCSB).  One of its corollaries is the emergence of pseudoscalar Nambu--Goldstone bosons---most recognisably, the~pions; and their appearance and properties have a profound impact on the character of the physical Universe.  Nuclear physics, e.g.,\ the~formation of the elements and the number of elements that may be formed, seems to be very sensitive to the pion mass; and the pion mass is a quantity which is uniquely sensitive to the interplay between emergent and explicit (Higgs) mass generating~mechanisms.

A non-perturbative framework that can unify the emergence of gluon and quark masses, and~express their impacts on solutions of the associated bound-state equations in quantum field theory and also in the matrix elements which describe complex observable processes, can provide reductive explanations for physical phenomena.   No single such framework exists, but~an amalgam of high-level approaches is bearing~fruit.




\vspace{6pt}

\funding{This work was supported in part by Jiangsu Province Hundred Talents Plan for Professionals.}

%


\acknowledgments{These notes are based on results obtained and insights developed through collaborations with many people, to~all of whom I am greatly~indebted.}

\conflictsofinterest{The author declares no conflicts of~interest.}

\newpage
\abbreviations{The following abbreviations are used in this manuscript.\\

\noindent
\begin{tabular}{ll}
CSM & continuum Schwinger-function method \\
DA (PDA) & (parton) distribution amplitude \\
DF (PDF) & (parton) distribution function \\
DCSB & dynamical chiral symmetry breaking \\
DSE & Dyson--Schwinger Equation\\
EFT  & effective field theory\\
EHM & emergent hadronic mass\\
ERBL & Efremov--Radyushkin--Brodsky--Lepage \\
JLab & Thomas Jefferson National Accelerator Facility\\
lQCD & lattice-regularised quantum chromodynamics\\
PDG & Particle Data Group \\
PI & process independent \\
PFF & parton fragmentation function\\
pQCD & perturbative quantum chromodynamics\\
QED & quantum electrodynamics\\
QCD & quantum chromodynamics \\
RGI & renormalisation group invariant\\
SM & Standard Model (of Particle Physics)\\
SPM & Schlessinger point method \\
2PI & two-particle irreducible
\end{tabular}}

\reftitle{References}

\end{document}